\documentclass[12pt]{article}

\usepackage{amsmath,amssymb,amsfonts,amsthm}
\usepackage[T1,T2A]{fontenc}
\usepackage{graphicx}
\usepackage{cite}
\usepackage[all]{xy}
\usepackage[toc,page]{appendix}
\usepackage{hyperref}
\hypersetup{colorlinks=true,  citecolor=red, linkcolor=blue}
\usepackage[mathcal]{eucal}

\allowdisplaybreaks[3]

\newcounter{propositiona}

\newcounter{definitiona}

\newcounter{remarka}
\newcommand{\remarka}[1]{\refstepcounter{remarka}
\noindent
\textbf{Remark \theremarka.}\, #1}
\newcounter{examplea}

\newcounter{lemmaa}
\newcommand{\lemmaa}[1]{\refstepcounter{lemmaa}
\noindent
\textbf{Lemma \thelemmaa.}\, {\it #1}}
\newcounter{theorema}
\newcommand{\theorema}[1]{\refstepcounter{theorema}
\noindent
\textbf{Theorem\, \thetheorema.}\, {\it #1}}
\newcounter{corollarya}

\renewcommand{\thefootnote}{\alph{footnote}}

\title{Invariant Reduction for Partial Differential Equations. IV: Symmetries that Rescale Geometric Structures} 

\author{ \renewcommand{\thefootnote}{\alph{footnote}}
Kostya Druzhkov\hspace{0.1ex}\footnotemark[1],~~Alexei Cheviakov\footnotemark[2]\vspace{0.5cm}\\
\small \emph{Department of Mathematics and Statistics, University of Saskatchewan, Saskatoon, Canada}\vspace{0.2cm}\\
}

\begin{document}

\footnotetext[1]{Electronic mail: konstantin.druzhkov@gmail.com}
\footnotetext[2]{Electronic mail: shevyakov@math.usask.ca}

\maketitle \numberwithin{equation}{section}
\renewcommand{\thefootnote}{\arabic{footnote}}

\def\beq{\begin{equation}}
\def\eeq{\end{equation}}
\def\barr{\begin{array}{ll}}
\def\earr{\end{array}}

\begin{abstract}
For a system of partial differential equations admitting point, contact, or higher symmetries,
the framework of invariant reduction systematically computes how invariant
geometric structures, such as conservation laws, presymplectic structures,
variational principles, and Poisson brackets, are inherited by the systems
governing symmetry-invariant solutions. We extend this mechanism to geometric structures that are not invariant but are \emph{rescaled} by a symmetry.
Specifically, if $X$ is the symmetry used for reduction, $X_s$ is
a symmetry satisfying $[X_s,X]=aX$, and the Lie derivative
$\mathcal{L}_{X_s}$ acts on an $X$-invariant element of the
Vinogradov $\mathcal{C}$-spectral sequence as multiplication by~$b$, then the
restricted symmetry $X_s|_{\mathcal{E}_X}$ acts on the corresponding
reduction as multiplication by~$a+b$.
This shift rule gives rise to two phenomena:
the \emph{emergence of invariance}, where reductions acquire an
invariance that was not present at the level of the original structure,
and the \emph{loss of invariance}, where reductions of invariant
structures are no longer invariant.
As an application, we describe a class of exact solutions to systems
possessing sufficiently many symmetries and conservation laws subject
to certain compatibility conditions.
These solutions are invariant under pairs of symmetries and are
completely determined by explicitly constructed functions that are
constant on them; the description is geometric and does not require any integrability-related structures such as Lax pairs.
The framework is illustrated by two examples:
the Lin--Reissner--Tsien equation of potential nonstationary transonic
gas flows, for which closed-form exact solutions are obtained and
validated numerically, and the potential Boussinesq system, for which
the inherited Poisson bracket is employed to describe solutions
determined by algebraic equations.
\end{abstract}


\section{Introduction}

Invariant (symmetry) reduction is one of the principal methods for constructing and analyzing
special solution classes of nonlinear partial differential equations (PDEs). Given a system $\mathcal E$ and a
symmetry $X$, the $X$-invariant solutions are described by the overdetermined system obtained by
augmenting $\mathcal E$ with the invariant surface condition given by the relation $\varphi=0$ and its differential consequences, where $\varphi$ is the characteristic of $X$. In $(1+1)$-dimensional PDEs, such reductions under point symmetries commonly lead to ordinary differential equation (ODE) systems whose solutions can be analyzed using tools familiar from ordinary differential equations.

In Part~I \cite{druzhkov2025invariant} of this series, we used invariant conservation laws to produce \emph{constants of $X$-invariant motion} in the $(1+1)$-dimensional setting, without introducing special coordinates and
without requiring that $X$ generates a flow. The intrinsic formulation identifies
conservation laws with cohomology classes in Vinogradov's $\mathcal{C}$-spectral sequence and interprets the
reduction mechanism as a descent in the bidegree of the sequence. Subsequent parts extend the
framework to the general multidimensional mechanism applicable to cohomological structures that can be defined intrinsically (Part~II \cite{druzhkovcheviakov2025invariantreductionpartialdifferential}) and to Poisson
structures (Part~III \cite{druzhkov2025invariantreductionpartialdifferential}).

The present paper addresses a phenomenon that frequently appears in symmetry reductions of physically motivated
PDEs: even when one reduces with respect to
a symmetry $X$, the reduced system $\mathcal E_X$ may inherit additional symmetries of the original
equation that are \emph{not} $X$-invariant. A typical and important case is a \emph{scaling symmetry}
$X_s$ that rescales the reduction-generating symmetry,
\begin{equation}\label{eq:comm-rel-intro}
[X_s,X]=aX,\qquad a\in\mathbb R,
\end{equation}
and simultaneously rescales geometric objects such as conservation laws, variational forms, or
other cohomology classes. Understanding how such rescalings interact with invariant reduction is
crucial for (i) determining how reduced structures interact on $\mathcal E_X$, (ii) identifying when
invariance \emph{emerges} after reduction, (iii) extending the straightforward multi-reduction procedure described in Part II~\cite{druzhkovcheviakov2025invariantreductionpartialdifferential}, and (iv) explaining situations where invariance is
\emph{lost} although it is present before reduction.

Our first contribution is a general theorem describing the induced action of $X_s$ on reductions of
$X$-invariant cohomology classes in the $\mathcal{C}$-spectral sequence. Let $\omega\in E^{p,\hspace{0.15ex} q}_0(\mathcal E)$
represent an $X$-invariant element of $E^{p,\hspace{0.15ex} q}_1(\mathcal E)$, so that there exists
$\vartheta\in E^{p,\hspace{0.15ex} q-1}_0(\mathcal E)$ with
\begin{equation}\label{eq:basic-reduction-intro}
\mathcal{L}_X\omega=d_0\vartheta.
\end{equation}
The restriction $\vartheta|_{\mathcal E_X}$ represents the invariant reduction of the class. Suppose, moreover, that $X_s$ acts on the cohomology class $[\omega]\in E^{p,\hspace{0.15ex} q}_1(\mathcal E)$ of
$\omega$ by scaling,
\begin{equation}\label{eq:scaling-intro}
\mathcal{L}_{X_s}[\omega]=b\hspace{0.15ex} [\omega],
\end{equation}
for some $b\in\mathbb R$. We show that, under
mild well-definedness assumptions, the induced Lie derivative along the restricted symmetry
$\widetilde X_s=X_s|_{\mathcal E_X}$ acts on the reduced class as multiplication by $a+b$, where $a$ is
from~\eqref{eq:comm-rel-intro}. In particular, when $a+b=0$, the reduction becomes
$\widetilde X_s$-invariant even if $b\neq 0$, providing a precise mechanism for the \emph{emergence of
invariance}. Conversely, if $a\neq 0$ and $b=0$, a nontrivial reduction fails to remain invariant,
which we interpret as a mechanism for the \emph{loss of invariance} after reduction.

The second contribution concerns exact solutions. The above rescaling mechanism implies that a
finite-dimensional invariant manifold $\mathcal E_X$ may possess subsystems defined by vanishing of scalar invariants
arising in reduction (reduced invariants) that inherit both commuting symmetries and a scaling symmetry. Under natural
algebraic and regularity assumptions, this leads to a description of a class of solutions that are
simultaneously invariant under pairs of symmetries of $\mathcal E$, namely under $X$ and
$X_s- C^i X_i$, where the $X_i$ commute with $X$ and the cosets $X_i + \langle X \rangle$ transform linearly under $X_s$.
Geometrically, the constants $C^i$ arise as first integrals on the finite-dimensional subsystem of $\mathcal{E}_X$ and
are constant along its solutions.

We illustrate the theory in two examples. First, we consider the Lin--Reissner--Tsien equation for
potential nonstationary transonic gas flows. We exhibit a conservation law belonging to a family
parameterized by an arbitrary function of one independent variable and show that its $X$-reduction
acquires scaling invariance. We then derive explicit families of solutions invariant under both $X$
and a suitably chosen $X_s$, given by quadrature via the integral of an explicitly constructed closed $1$-form.
We also validate a representative exact solution numerically by evolving the associated first-order
system for its derivatives using a WENO5--SSPRK(3,3) scheme and a projection
step enforcing the integrability constraint. Second, we study the potential Boussinesq system in
the Hamiltonian setting: using a local Hamiltonian operator and several $X$-invariant conservation
laws, we construct multiple reduced invariants and obtain a finite-dimensional subsystem whose
solutions are characterized by explicit algebraic relations together with dynamically selected
symmetry constraints.

The paper is organized as follows. Section~\ref{Basnot} reviews the required geometric background on jets,
the Cartan distribution, and the $\mathcal{C}$-spectral sequence. Section~\ref{Mainsection} formulates invariant reduction at
the level of $E^{\hspace{0.1ex} p,\hspace{0.15ex} q}_1(\mathcal E)$ and proves the main theorem on the action of rescaling symmetries
on reduced classes. Section~\ref{SectionScalsymexectsols} derives the symmetry-based description of the corresponding class of
exact solutions. Section~\ref{Sectionexamples} presents the examples and, in the first example, the accompanying
numerical validation. An appendix contains the proof of an auxiliary lemma used in the geometric
construction. Throughout the paper, all functions are assumed to be smooth of the class $C^{\infty}$, and all manifolds are assumed to have no boundary.

\section{Basic notation and definitions \label{Basnot}}

Let us introduce notation and briefly recall basic facts from the geometry of differential equations.

\subsection{Jets}

We now briefly review the notion of jet bundles and related structures. For details, see, e.g.,~\cite{VinKr}.

\vspace{1ex}

Let $\pi\colon E^{n+m}\to M^n$ be a locally trivial smooth vector bundle over a smooth manifold $M^n$. Suppose $U\subset M$ is a coordinate neighborhood such that the bundle $\pi$
becomes trivial over $U$. Choose local coordinates $x^1$, \ldots, $x^n$ in $U$ and $u^1$, \ldots, $u^m$
along the fibers of $\pi$ over $U$. In these coordinates, a section $\sigma\in \Gamma(\pi)$ has the form of a smooth vector function
\begin{align*}
\sigma\colon\qquad u^1 = \sigma^1(x^1, \ldots, x^n)\,,\qquad \ldots\,, \qquad u^m = \sigma^m(x^1, \ldots, x^n)\,.
\end{align*}
Two sections $\sigma_1, \sigma_2 \in \Gamma(\pi)$ define the same $k$-jet ($k = 0, 1, \ldots, \infty$) at a point $x_0\in U$, if for $i = 1, \ldots, m$, the functions $\sigma^i_1$ and $\sigma^i_2$ have the same $k$-degree Taylor polynomials at $x_0$. We denote by $[\sigma]^k_{x_0}$ the $k$-jet of $\sigma\in \Gamma(\pi)$ at $x_0\in M$. The set $J^{k}(\pi)$ of all $k$-jets of sections of $\pi$ is naturally endowed with a smooth manifold structure.
On $J^{k}(\pi)$, one can introduce adapted local coordinates, given by
\begin{align}
x^i([\sigma]^{k}_{x_0}) = x^i_0\,,\qquad u^i_{\alpha}([\sigma]^{k}_{x_0}) = \dfrac{\partial^{|\alpha|} \sigma^i}{(\partial x^1)^{\alpha_1}\ldots (\partial x^n)^{\alpha_n}}(x_0)\,,\qquad |\alpha| \leqslant k\,.
\label{adaptedcoordinates}
\end{align}
Here $\alpha$ is a multi-index, $|\alpha| = \alpha_1 + \ldots + \alpha_n$.
It is convenient to treat $\alpha$ as a formal sum of the form $\alpha = \alpha_1 x^1 + \ldots + \alpha_n x^n = \alpha_i x^i$, where all $\alpha_i$ are non-negative integers. In what follows, we consider only adapted local coordinates on $J^{k}(\pi)$.


\vspace{0.5ex}
\noindent
\textbf{Functions.}
The projections $\pi_{\infty, \,k}\colon J^{\infty}(\pi)\to J^k(\pi)$, $[\sigma]^{\infty}_{x_0}\mapsto [\sigma]^{k}_{x_0}$ allow one to define the algebra (over $\mathbb{R}$) of smooth functions on the infinite jet space $J^{\infty}(\pi)$
$$
\mathcal{F}(\pi) = \bigcup_{k\geqslant 0} \pi_{\infty,\hspace{0.2ex} k}^{\hspace{0.1ex} *}\, C^{\infty}(J^k(\pi))\,.
$$

\noindent
\textbf{Cartan distribution.} The main structure on jet manifolds is the Cartan distribution.
Using adapted local coordinates on $J^{\infty}(\pi)$, one can introduce the total derivatives
$$
D_{x^i} = \partial_{x^i} + u^k_{\alpha + x^i}\partial_{u^k_{\alpha}}\qquad\quad i = 1, \ldots, n.
$$
The planes of the Cartan distribution $\mathcal{C}$ on $J^{\infty}(\pi)$ are spanned by the total derivatives. It is convenient to interpret tangent vectors/vector fields on $J^{\infty}(\pi)$ in terms of derivations of $\mathcal{F}(\pi)$.

\vspace{0.5ex}
\noindent
\textbf{Cartan forms.} The Cartan distribution $\mathcal{C}$ determines the ideal $\mathcal{C}\Lambda^*(\pi)$
of the algebra
$$
\Lambda^*(\pi) = \bigcup_{k\geqslant 0} \pi_{\infty,\hspace{0.2ex} k}^{\hspace{0.1ex} *}\, \Lambda^*(J^k(\pi))
$$
of differential forms on $J^{\infty}(\pi)$.
The ideal $\mathcal{C}\Lambda^*(\pi)$ is generated by Cartan (or contact) forms, i.e., differential forms that vanish on all planes of the Cartan distribution $\mathcal{C}$.
A Cartan $1$-form $\omega\in\mathcal{C}\Lambda^1(\pi)$ can be written as a finite sum
$$
\omega = \omega_i^{\alpha}\theta^i_{\alpha}\,,\qquad\ \theta^i_{\alpha} = du^i_{\alpha} - u^i_{\alpha + x^k}dx^k
$$
in adapted local coordinates. The coefficients $\omega_i^{\alpha}$ are smooth functions of adapted coordinates.

\vspace{0.5ex}
\noindent
\textbf{Infinitesimal symmetries.} Denote by $\pi_k$ the projection $\pi_k\colon J^k(\pi)\to M$, $\pi_k\colon [\sigma]^k_{x_0}\mapsto x_0$.
Smooth sections of the pullback bundles $\pi^*_{k}(\pi)\colon \pi^*_{k}(E)\to J^k(\pi)$ naturally determine sections of the pullback $\pi_{\infty}^*(\pi)\colon \pi^*_{\infty}(E)\to J^{\infty}(\pi)$ by means of the projections $\pi_{\infty,\hspace{0.2ex} k}$. We denote by $\varkappa(\pi)$ the $\mathcal{F}(\pi)$-module of such sections of $\pi_{\infty}^*(\pi)$. Each section $\varphi\in \varkappa(\pi)$ gives rise to a unique evolutionary vector field $E_{\varphi}$ on $J^{\infty}(\pi)$. In adapted coordinates,
$$
E_{\varphi} = D_{\alpha}(\varphi^i)\partial_{u^i_{\alpha}}\,,
$$
where $\varphi^1$, \ldots, $\varphi^m$ are components of $\varphi$, $D_{\alpha}$ denotes the composition $D_{x^1}^{\ \alpha_1}\circ\ldots\circ D_{x^n}^{\ \alpha_n}$.
Evolutionary vector fields are evolutionary symmetries of $J^{\infty}(\pi)$. In particular, $\mathcal{L}_{E_{\varphi}}\,\mathcal{C}\Lambda^*(\pi)\subset \mathcal{C}\Lambda^*(\pi)$. Here $\mathcal{L}_{E_{\varphi}}$ is the corresponding Lie derivative. Elements of $\varkappa(\pi)$ are characteristics of symmetries of~$J^{\infty}(\pi)$. An infinitesimal symmetry of $J^{\infty}(\pi)$ is a sum of an evolutionary vector field and a trivial symmetry, i.e., a combination of the total derivatives $D_{x^1}$, $\ldots$, $D_{x^n}$.

\vspace{0.5ex}
\noindent
\textbf{Horizontal forms.}
Cartan forms allow one to consider the modules of horizontal $k$-forms
$$
\Lambda^k_h(\pi) = \Lambda^k(\pi)/\mathcal{C}\Lambda^k(\pi)\,.
$$
The de Rham differential $d$ induces the horizontal differential $d_h\colon \Lambda^k_h(\pi)\to \Lambda^{k+1}_h(\pi)$.

\subsection{Differential equations \label{SectionDiffEq}}

Let $\zeta \colon E_1\to M$ be a locally trivial smooth vector bundle over the same base as $\pi$. Smooth sections of the pullbacks $\pi^*_{r}(\zeta)$ determine a module of sections of the pullback $\pi^*_{\infty}(\zeta)\colon \pi^*_{\infty}(E_1)\to J^{\infty}(\pi)$. We denote it by $P(\pi)$. Any $F\in P(\pi)$ can be considered a (generally, nonlinear) differential operator $\Gamma(\pi)\to \Gamma(\zeta)$. Then $F = 0$ is a differential equation.
By its \emph{infinite prolongation} we mean the set of formal solutions $\mathcal{E}\subset J^{\infty}(\pi)$ defined by the infinite system of equations
\begin{align*}
\mathcal{E}\colon\qquad D_{\alpha}(F^i) = 0\,,\qquad |\alpha| \geqslant 0\,.
\end{align*}
Here $F^i$ are components of $F$ in adapted coordinates. We denote $\pi_{\mathcal{E}} = \pi_{\infty}|_{\mathcal{E}}$ and assume that $\pi_{\mathcal{E}}\colon \mathcal{E}\to M$ is surjective.

\vspace{1ex}
\remarka{We do not require that the number of equations of the form $F^i = 0$ coincide with the number of dependent variables $m$.}

\vspace{1ex}
\noindent
\textbf{Regularity assumptions.} We say that $\mathcal{E}\subset J^{\infty}(\pi)$ is \emph{regular} if
\begin{enumerate}
  \item There exists $r\in \mathbb{Z}$ such that for all $k\geqslant r$, the images $\mathcal{E}_k = \pi_{\infty,\hspace{0.1ex} k}(\mathcal{E})\subset J^{k}(\pi)$ are properly embedded submanifolds, and the projections $\pi_{k+1,\hspace{0.1ex} k}|_{\mathcal{E}_{k+1}}\colon \mathcal{E}_{k+1}\to \mathcal{E}_{k}$ are submersions.
  \item A function $f\in \mathcal{F}(\pi)$ vanishes on $\mathcal{E}$ if and only if there is a differential operator $\Delta\colon P(\pi)\to \mathcal{F}(\pi)$ of the form $\Delta_i^{\alpha} D_{\alpha}$ (total differential operator) such that $f = \Delta(F)$. Here, for some integer $k$, the components $\Delta_i^{\alpha}$ with $|\alpha|\leqslant k$ may depend on the independent variables $x^i$, the dependent variables $u^i$, and derivatives up to order $k$, while $\Delta_i^{\alpha} = 0$ for $|\alpha| > k$.
\end{enumerate}


Unless otherwise stated, we assume that all systems considered below are regular. We also assume that the de Rham cohomology groups $H^i_{dR}$ of all considered systems are trivial for $i > 0$.

\vspace{0.5ex}
\noindent
\textbf{Functions.}
By $\mathcal{F}(\mathcal{E})$ we denote the algebra of smooth functions on $\mathcal{E}$,
$$
\mathcal{F}(\mathcal{E}) = \mathcal{F}(\pi)|_{\mathcal{E}} = \mathcal{F}(\pi)/I\,.
$$
Here $I$ denotes the ideal of the system $\mathcal{E}\subset J^{\infty}(\pi)$, $I = \{f\in \mathcal{F}(\pi)\, \colon\, f|_{\mathcal{E}} = 0\}$. Tangent vectors/vector fields on $\mathcal{E}$ are defined in terms of derivations of the algebra $\mathcal{F}(\mathcal{E})$.

\vspace{0.5ex}
\noindent
\textbf{Cartan forms.} The ideal $\mathcal{C}\Lambda^*(\pi)\subset \Lambda^*(\pi)$ gives rise to the corresponding ideal $\mathcal{C}\Lambda^*(\mathcal{E})$ of the algebra $\Lambda^*(\mathcal{E}) = \Lambda^*(\pi)/(I\cdot \Lambda^*(\pi) + dI\wedge \Lambda^*(\pi))$ of differential forms on $\mathcal{E}$. The Cartan plane at a point of $\mathcal{E}$ is spanned by tangent vectors that annihilate all Cartan $1$-forms.

\vspace{0.5ex}
\noindent
\textbf{Solutions.} A (local) section $\sigma$ of $\pi_{\mathcal{E}}\colon \mathcal{E}\to M$ is a (local) \emph{solution} of the equation $\pi_{\mathcal{E}}$ if its differential maps tangent planes to Cartan planes. Equivalently, if $\sigma^*(\mathcal{C}\Lambda^1(\mathcal{E}))=0$. 
We consider only smooth sections of $\pi_{\mathcal{E}}$, i.e., such that $\sigma^*(\mathcal{F}(\mathcal{E}))\subset C^{\infty}(M)$. Smooth (maximal) integral manifolds of the Cartan distribution of $\mathcal{E}$ play the role of (global) solutions of $\mathcal{E}$.

\vspace{0.5ex}
\noindent
\textbf{Infinitesimal symmetries.} A \emph{symmetry} (more precisely, an infinitesimal symmetry) of an infinitely prolonged system of equations $\mathcal{E}$ is a vector field $X$ on $\mathcal{E}$ (a derivation of $\mathcal{F}(\mathcal{E})$) that preserves the Cartan distribution: $[X, \mathcal{C}D(\mathcal{E})]\subset \mathcal{C}D(\mathcal{E})$, where $\mathcal{C}D(\mathcal{E})$ denotes the module of \emph{Cartan derivations}, i.e., vector fields on $\mathcal{E}$ whose vectors lie in the respective planes of the Cartan distribution. Two symmetries are equivalent if they differ by a Cartan derivation (a trivial symmetry). One can say that, locally, trivial symmetries of a regular system are combinations of the total derivatives $\,\overline{\!D}_{x^i} = D_{x^i}|_{\mathcal{E}}$, $i = 1, \ldots, n$.

If $\varphi\in \varkappa(\pi)$ is a characteristic such that $E_{\varphi}$ is tangent to $\mathcal{E}$ (i.e., $E_{\varphi}(I)\subset I$), then the restriction $E_{\varphi}|_{\mathcal{E}}\colon \mathcal{F}(\mathcal{E})\to \mathcal{F}(\mathcal{E})$ is an evolutionary symmetry of $\mathcal{E}$ (less formally, $E_{\varphi}$ can also be called a symmetry of $\mathcal{E}$).
Evolutionary symmetries of a regular system $\mathcal{E}\subset J^{\infty}(\pi)$ are in one-to-one correspondence with elements of the kernel of the linearization operator $l_{\mathcal{E}} = l_F|_{\mathcal{E}}\colon \varkappa(\mathcal{E})\to P(\mathcal{E})$, where $l_F\colon \varkappa(\pi)\to P(\pi)$, $\varphi \mapsto E_{\varphi}(F)$, $l_F(\varphi)^i = E_{\varphi}(F^i)$, and
$$
\varkappa(\mathcal{E}) = \varkappa(\pi)/I\cdot \varkappa(\pi)\,,\qquad P(\mathcal{E}) = P(\pi)/I\cdot P(\pi)\,.
$$

\vspace{0.5ex}
\noindent
\textbf{$\mathcal{C}$-spectral sequence.} For $p \geqslant 1$, the $p$-th power $\mathcal{C}^p\Lambda^*(\mathcal{E})$ of the ideal $\mathcal{C}\Lambda^*(\mathcal{E})$ is stable with respect to the de Rham differential $d$, i.e., $d(\mathcal{C}^p\Lambda^*(\mathcal{E})) \subset \mathcal{C}^p\Lambda^*(\mathcal{E})$.
Then the de Rham complex $\Lambda^{\bullet}(\mathcal{E})$ admits the filtration
$$
\Lambda^{\bullet}(\mathcal{E})\supset \mathcal{C}\Lambda^{\bullet}(\mathcal{E})\supset \mathcal{C}^2\Lambda^{\bullet}(\mathcal{E})\supset \mathcal{C}^3\Lambda^{\bullet}(\mathcal{E})\supset \ldots
$$
The corresponding spectral sequence $(E^{\hspace{0.1ex} p,\hspace{0.2ex} q}_r(\mathcal{E}), d^{\hspace{0.1ex} p,\hspace{0.2ex} q}_r)$ is the Vinogradov $\mathcal{C}$-spectral sequence~\cite{VINOGRADOV198441, VinKr}.
Here $\mathcal{C}^{k+1}\Lambda^k(\mathcal{E}) = 0$, $E^{\hspace{0.1ex} p, \hspace{0.2ex} q}_0(\mathcal{E}) = \mathcal{C}^p\Lambda^{p+q}(\mathcal{E})/\mathcal{C}^{p+1}\Lambda^{p+q}(\mathcal{E})$. All differentials $d_r^{\hspace{0.1ex} p,\hspace{0.2ex} q}$ are induced by the de Rham differential $d$. In this paper, we consider only $d_0^{\hspace{0.1ex} p,\hspace{0.2ex} q}\colon E^{\hspace{0.1ex} p, \hspace{0.2ex} q}_0(\mathcal{E}) \to E^{\hspace{0.1ex} p, \hspace{0.2ex} q+1}_0(\mathcal{E})$ and
$$
d_1^{\hspace{0.1ex} p,\hspace{0.2ex} q}\colon E^{\hspace{0.1ex} p, \hspace{0.2ex} q}_1(\mathcal{E}) \to E^{\hspace{0.1ex} p+1, \hspace{0.2ex} q}_1(\mathcal{E})\,,\qquad E^{\hspace{0.1ex} p, \hspace{0.2ex} q}_{1}(\mathcal{E}) = \ker d_0^{\hspace{0.1ex} p,\hspace{0.2ex} q}/ \mathrm{im}\, d_0^{\hspace{0.1ex} p,\hspace{0.2ex} q-1}\,.
$$
We also use the notation $d_0$, $d_1$ where it does not lead to confusion.


In the regular case, each element of $E^{\hspace{0.1ex} p, \hspace{0.2ex} q}_0(\mathcal{E})$ has a unique representative in the restriction of $\mathcal{C}^p\Lambda^p(\pi)\wedge \pi_{\infty}^*(\Lambda^q(M))$.
To make the description less abstract, we identify elements of $E^{\hspace{0.1ex} p, \hspace{0.2ex} q}_0(\mathcal{E})$ with their representatives of this form. Locally, on a regular $\mathcal{E}$, the differential $d_0$ takes the form
$$
d_0 = dx^i\wedge \mathcal{L}_{\,\overline{\!D}_{x^i}}\,,\qquad \,\overline{\!D}_{x^i} = D_{x^i}|_{\mathcal{E}}\,.
$$

\vspace{0.5ex}

\noindent
\textbf{Conservation laws.} A \emph{conservation law} of $\mathcal{E}$ is an element of the group $E^{\hspace{0.1ex} 0,\hspace{0.2ex} n-1}_1(\mathcal{E})$. Let us introduce the modules
\begin{align*}
\widehat{P}(\pi) = \mathrm{Hom}_{\mathcal{F}(\pi)}(P(\pi), \Lambda^n_h(\pi))\,,\qquad \widehat{\varkappa}(\pi) = \mathrm{Hom}_{\mathcal{F}(\pi)}(\varkappa(\pi), \Lambda^n_h(\pi))\,.
\end{align*}
A \emph{characteristic} of a conservation law is any homomorphism $\psi\in \widehat{P}(\pi)$ such that
\begin{align*}
\langle \psi, F \rangle = d_h\hspace{0.15ex} \mu\,,\qquad \mu\in\Lambda_h^{n-1}(\pi)\,,
\end{align*}
where $\mu|_{\mathcal{E}}\in\Lambda_h^{n-1}(\mathcal{E}) = E_0^{\hspace{0.1ex} 0,\hspace{0.2ex} n-1}(\mathcal{E})$ represents the conservation law.

\vspace{0.5ex}

\noindent
\textbf{Cosymmetries.} Elements of the kernel of the adjoint linearization operator $l_{\mathcal{E}}^{\, *}\colon \widehat{P}(\mathcal{E})\to \widehat{\varkappa}(\mathcal{E})$ are \emph{cosymmetries} of $\mathcal{E}$, provided $\mathcal{E}$ is regular. Here $l_{\mathcal{E}}^{\, *} = l_{F}^{\, *}|_{\mathcal{E}}$, $\widehat{P}(\mathcal{E}) = \widehat{P}(\pi)/I \cdot \widehat{P}(\pi)$, $\widehat{\varkappa}(\mathcal{E}) = \widehat{\varkappa}(\pi)/I \cdot \widehat{\varkappa}(\pi)$. The restriction $\,\overline{\!\psi} = \psi|_{\mathcal{E}}$ of a characteristic $\psi$ of a conservation law is a cosymmetry.

\section{Invariant reduction and rescaling of geometric structures \label{Mainsection}}

Let us briefly recall the invariant reduction framework ~\cite{druzhkovcheviakov2025invariantreductionpartialdifferential}. Let $\mathcal{E}\subset J^{\infty}(\pi)$ be the infinite prolongation of a system of differential equations $F = 0$. If $X = E_{\varphi}|_{\mathcal{E}}$ is a symmetry of $\mathcal{E}$, $\varphi\in \varkappa(\pi)$, then $X$-invariant solutions of $\mathcal{E}$ are described by the infinite prolongation $\mathcal{E}_X$ of the system
\begin{align*}
F = 0\,,\qquad \varphi = 0\,.
\end{align*}
Suppose $\omega\in E^{\hspace{0.1ex} p, \hspace{0.2ex} q}_0(\mathcal{E})$ represents an $X$-invariant element of a group $E^{\hspace{0.1ex} p, \hspace{0.2ex} q}_1(\mathcal{E})$ of Vinogradov's $\mathcal{C}$-spectral sequence. Then there exists $\vartheta\in E^{\hspace{0.1ex} p, \hspace{0.2ex} q-1}_0(\mathcal{E})$ such that
\begin{align}
\mathcal{L}_X \omega = d_0 \vartheta.
\label{redformula}
\end{align}
The system $\mathcal{E}_X\subset \mathcal{E}$ is characterized by the condition that $X$ vanishes at its points. Then~\eqref{redformula} implies that
\begin{align*}
d_0 \vartheta|_{\mathcal{E}_X} = 0\,,
\end{align*}
and $\vartheta|_{\mathcal{E}_X}\in E^{\hspace{0.1ex} p, \hspace{0.2ex} q-1}_0(\mathcal{E}_X)$ represents an element of $E^{\hspace{0.1ex} p, \hspace{0.2ex} q-1}_1(\mathcal{E}_X)$. This mapping from $X$-invariant elements of $E^{\hspace{0.1ex} p, \hspace{0.2ex} q}_1(\mathcal{E})$ to $E^{\hspace{0.1ex} p, \hspace{0.2ex} q-1}_1(\mathcal{E}_X)$ is well-defined if and only if $E^{\hspace{0.1ex} p, \hspace{0.2ex} q-1}_1(\mathcal{E})|_{\mathcal{E}_X} = 0$. We also say that it is well-defined in the case $p = 0$, $q = 1$, $E^{\hspace{0.1ex} 0, \hspace{0.2ex} 0}_1(\mathcal{E})|_{\mathcal{E}_X} \subset H^0_{dR}(\mathcal{E}_X)$, as we consider reductions of elements of the group $E^{\hspace{0.1ex} 0, \hspace{0.2ex} 1}_1(\mathcal{E})$ modulo additive locally constant functions on $\mathcal{E}_X$. In other words, the invariant reduction of $X$-invariant elements of $E^{\hspace{0.1ex} 0, \hspace{0.2ex} 1}_1(\mathcal{E})$ yields elements\footnote{This situation is similar to the reduction of invariant elements in other groups; recall that, for simplicity, we restrict ourselves to $\mathcal{E}_X$ such that $H^i_{dR}(\mathcal{E}_X) = 0$ for $i > 0$.} of $E^{\hspace{0.1ex} 0, \hspace{0.2ex} 0}_1(\mathcal{E}_X)/H^0_{dR}(\mathcal{E}_X)$.

The system $\mathcal{E}_X$ may inherit symmetries of $\mathcal{E}$ even if they are not $X$-invariant. A typical example is provided by scaling symmetries, which often rescale various geometric structures. In the context of the invariant reduction, this situation is described by the following theorem.

\vspace{1ex}

\theorema{\label{Theorem1} Let $X_s$ be an evolutionary symmetry of $\mathcal{E}$ satisfying $[X_s, X] = aX$ for some $a\in \mathbb{R}$. Suppose that the invariant reduction of $X$-invariant elements of $E^{\hspace{0.1ex} p, \hspace{0.2ex} q}_1(\mathcal{E})$ is well-defined. If $\omega\in E^{\hspace{0.1ex} p, \hspace{0.2ex} q}_0(\mathcal{E})$ represents an $X$-invariant element of $E^{\hspace{0.1ex} p, \hspace{0.2ex} q}_1(\mathcal{E})$ such that the Lie derivative $\mathcal{L}_{X_s}$ acts on this element as multiplication by $b\in \mathbb{R}$, then the Lie derivative along $X_s|_{\mathcal{E}_X}$ acts on its reduction as multiplication by $a + b$.
}
\vspace{0.5ex}

\noindent
\textbf{Proof.} There exist $\vartheta, \nu \in E^{\hspace{0.1ex} p, \hspace{0.2ex} q-1}_0(\mathcal{E})$ such that
\begin{align*}
\mathcal{L}_X \omega = d_0 \vartheta\,,\qquad \mathcal{L}_{X_s}\hspace{0.15ex} \omega = b\hspace{0.2ex} \omega + d_0 \nu\,.
\end{align*}
Applying $\mathcal{L}_X$ to the latter relation, one finds
\begin{align*}
\mathcal{L}_X \mathcal{L}_{X_s}\hspace{0.15ex} \omega = b\hspace{0.2ex} d_0 \vartheta + d_0 \mathcal{L}_X \nu\,.
\end{align*}
Taking into account the formula
$\mathcal{L}_X \mathcal{L}_{X_s} = \mathcal{L}_{X_s} \mathcal{L}_X + \mathcal{L}_{[X, X_s]} = \mathcal{L}_{X_s} \mathcal{L}_X - a\hspace{0.15ex} \mathcal{L}_{X}$, we find that
\begin{align*}
d_0 \mathcal{L}_{X_s} \vartheta - a\hspace{0.2ex} d_0 \vartheta = b\hspace{0.2ex} d_0 \vartheta + d_0 \mathcal{L}_X \nu\,.
\end{align*}
This relation can be rewritten as $d_0(\mathcal{L}_{X_s} \vartheta - a\vartheta - b\vartheta - \mathcal{L}_X \nu) = 0$. If $(p, q)\neq (0, 1)$, the restriction of $\mathcal{L}_{X_s} \vartheta - (a + b)\vartheta - \mathcal{L}_X \nu$ to $\mathcal{E}_X$ is $d_0$-exact. From $[X_s, X] = a X$ it follows that $\widetilde{X}_s = X_s|_{\mathcal{E}_X}$ is a symmetry of the system $\mathcal{E}_X$. Then on $\mathcal{E}_X$, we have
\begin{align*}
\mathcal{L}_{\widetilde{X}_s} (\vartheta|_{\mathcal{E}_X}) - (a + b) \vartheta|_{\mathcal{E}_X} \in \mathrm{im}\, d_0^{\hspace{0.15ex} p, \hspace{0.2ex} q-2}\,.
\end{align*}
Similarly, if $(p, q) = (0, 1)$, then
\begin{align}
\mathcal{L}_{\widetilde{X}_s} (\vartheta|_{\mathcal{E}_X}) - (a + b)\vartheta|_{\mathcal{E}_X} \in H^0_{dR}(\mathcal{E}_X)\,.
\label{invariantleaf}
\end{align}
This observation completes the proof.

\vspace{1ex}

\remarka{One can modify the symmetry $X_s$ by adding an evolutionary symmetry $X_0$ such that both the element of $E^{\hspace{0.1ex} p, \hspace{0.2ex} q}_1(\mathcal{E})$ and $X$ are $X_0$-invariant. \label{Rem1}}

\vspace{1ex}

A reformulation of Theorem \ref{Theorem1} for elements of the groups $\widetilde{E}^{\hspace{0.1ex} 0, \hspace{0.15ex} k}_1(\mathcal{E})$ of the spectral sequence for the Lagrangian formalism~\cite{DRUZHKOV2023104848} is straightforward. Let us recall that they encode variational principles~\cite{DRUZHKOV2024105143}. This may be seen as the formalism of Lagrangian multiforms~\cite{Suris2013, Suris2016} at the level of intrinsic geometry of PDEs. 

\vspace{1ex}

\remarka{It follows from~\eqref{invariantleaf} that if $H^0_{dR}(\mathcal{E}_X) = \mathbb{R}$, $(p, q) = (0, 1)$, and $a + b\neq 0$, then there exists a unique $C\in \mathbb{R}$ such that the subsystem of $\mathcal{E}_X$ defined by $\vartheta|_{\mathcal{E}_X} = C$ inherits the symmetry $\widetilde{X}_s$, provided that this subsystem is regular or that, for example, the vector field $\widetilde{X}_s$ generates a flow. If $(p, q) \neq (0, 1)$ and $a + b = 0$, then the reduction represented by $\vartheta|_{\mathcal{E}_X}$ is $\widetilde{X}_s$-invariant. When $b\neq 0$, this observation can be viewed as a mechanism for the emergence of invariance. \label{Rem2}}

\vspace{1ex}


Let us also note that Theorem~\ref{Theorem1} provides a mechanism for the loss of invariance when $a\neq 0$, $b = 0$, and the reduction represented by $\vartheta|_{\mathcal{E}_X}$ is nontrivial. This theorem admits a straightforward generalization to the case where the Lie derivative along $X_s$ of the element of $E^{\hspace{0.1ex} p, \hspace{0.2ex} q}_1(\mathcal{E})$ represented by $\omega$ is not necessarily proportional to it, but is still $X$-invariant. This generalization does not alter the mechanisms for the emergence or loss of invariance. Nonetheless, it allows for a slight extension of the approach in Section~\ref{SectionScalsymexectsols} (see Remark~\ref{Remongeneralization}).

\vspace{1ex}

\remarka{Theorem~\ref{Theorem1} clarifies how reductions of cotangent equations under two-dimensional algebras inherit their canonical variational $1$-forms in the $\ell$-normal case~\cite{druzhkov2025invariantreductionpartialdifferential}. Indeed, if $X$ and $X_s$ are evolutionary symmetries of $\mathcal{E}$ such that $[X_s, X] = aX$, then for their lifts\footnote{We use the notation from~\cite{druzhkov2025invariantreductionpartialdifferential}.} $\mathcal{X}_s$ and $\mathcal{X}$ to the cotangent equation $\mathcal{E}^*$, one has $[\mathcal{X}_s, \mathcal{X}] = a\mathcal{X}$. Since the canonical variational $1$-form $\rho\in E^{1,\hspace{0.1ex} n-1}_1(\mathcal{E}^*)$ is invariant under both $\mathcal{X}$ and $\mathcal{X}_s$, its reduction under $\mathcal{X}$ is invariant under the restriction of $\mathcal{X}_s$ if and only if $a = 0$. However, in accordance with Remark~\ref{Rem2}, the reduction of $\rho$ under $\mathcal{X}$ is invariant under the restriction of $\mathcal{X}_s - a\mathbb{X}$ to $\mathcal{E}^*_{\mathcal{X}}$ even if $a\neq 0$, where $\mathbb{X}$ denotes the canonical symmetry corresponding to $\rho$ (i.e., such that $\mathbb{X}\,\lrcorner\, d_1\rho = \rho$; in particular, $\mathbb{X}$ satisfies $[\mathbb{X}, \mathcal{X}] = 0$ and $\mathcal{L}_{\mathbb{X}}\hspace{0.15ex} \rho = \rho$).
The emergence of invariance occurs in both cases, whether the fibers of the natural projection $\mathcal{E}^*\to \mathcal{E}$ are odd or even. This observation is worth noting because invariant reduction of Hamiltonian operators is based on simultaneous reduction of the corresponding canonical variational $1$-forms.}

\section{Scaling symmetries and exact solutions \label{SectionScalsymexectsols}}

As a direct consequence of Theorem~\ref{Theorem1}, we outline a geometric\footnote{In particular, it does not require Lax pairs.} description of a class of solutions to systems possessing sufficiently many symmetries and conservation laws subject to certain conditions specified below. All such solutions are invariant under pairs of symmetries of $\mathcal{E}$ and are completely determined by explicitly constructed functions which are constant on them.
Although this description may be of limited practical use when the number of functions is large, the existence of solutions admitting such a description appears to be a structural feature of a class of $(1+1)$-dimensional integrable systems.\footnote{We emphasize that the current framework is applicable to general PDE systems, including non-integrable systems and overdetermined systems with more than two independent variables arising, for example, in reduction.}

More specifically, Remark~\ref{Rem2} shows that finite-dimensional systems of the form $\mathcal{E}_X$ may have subsystems that inherit $X$-invariant symmetries of $\mathcal{E}$ as well as its scaling symmetries. Sufficiently many independent constants of $X$-invariant motion arising from elements of $E^{\hspace{0.1ex} 0, \hspace{0.15ex} 1}_1(\mathcal{E})$ may lead to situations where solutions of the corresponding subsystems of $\mathcal{E}_X$ can be described through decomposing scaling symmetries into combinations of other symmetries with coefficients depending on solutions. This observation is apparently known in some form, but Theorem~\ref{Theorem1} and Remark~\ref{Rem2} provide a way to realize it as a systematic procedure.

Let $\mathcal{E}\subset J^{\infty}(\pi)$ be an infinitely prolonged system. We make the following assumptions.

\vspace{0.5ex}
\noindent
\textbf{1.} $X = E_{\varphi}|_{\mathcal{E}}$, $X_s = E_{\varphi_s}|_{\mathcal{E}}$, $X_1 = E_{\varphi_1}|_{\mathcal{E}}$, $\ldots$, $X_k = E_{\varphi_k}|_{\mathcal{E}}$ are symmetries of $\mathcal{E}$ such that $X$, $X_1$, $\ldots$, $X_k$ span a Lie algebra $\mathfrak{g}$, each $X_j$ commutes with $X$ and acts trivially on the quotient algebra $\mathfrak{g}/\langle X \rangle$ for $j = 1, \ldots, k$, while the Lie derivative $\mathcal{L}_{X_s}$ rescales $X$ and acts on $\mathfrak{g}/\langle X \rangle$ as an invertible linear operator, i.e.,
\begin{align}
[X_j, X] = 0\,,\qquad [X_i, X_j] = h_{ij} X,\qquad [X_s, X] = a X,\qquad [X_s, X_j] = c^i_j X_i + r_j X,
\label{actiononcommut}
\end{align}
where $h_{ij}, a, c^i_j, r_j\in \mathbb{R}$ and the matrix $(c^i_j)$ is invertible.\\
\textbf{2.} The corresponding system $\mathcal{E}_X$ is a connected finite-dimensional smooth manifold. Let us denote its dimension by $k + l + n$.\\
\textbf{3.} $\omega^1, \ldots, \omega^l\in E^{\hspace{0.1ex} 0, \hspace{0.2ex} 1}_0(\mathcal{E})$ represent $\mathfrak{g}$-invariant elements\footnote{By $\mathfrak{g}$-invariance we mean invariance under all elements of $\mathfrak{g}$.} of the group $E^{\hspace{0.1ex} 0, \hspace{0.2ex} 1}_1(\mathcal{E})$, for which the reduction under $X$ is well-defined, while the Lie derivative $\mathcal{L}_{X_s}$ acts on these elements of $E^{\hspace{0.1ex} 0, \hspace{0.2ex} 1}_1(\mathcal{E})$ as multiplication by real numbers different\footnote{One can allow them to coincide with $-a$ if, for example, $X_s$ vanishes at some point of $\mathcal{E}_X\subset \mathcal{E}$. In this case, Remark~\ref{Rem_allinvsolareontheleaf} ceases to be valid.} from $-a$.

\vspace{0.5ex}
\noindent
In this case, as follows from Remark~\ref{Rem2}, there exist functions $\vartheta^1, \ldots, \vartheta^l\in \mathcal{F}(\mathcal{E})$ such that
\begin{align*}
\mathcal{L}_{X} \omega^i = d_0 \vartheta^i
\end{align*}
and that for $I^i = \vartheta^i|_{\mathcal{E}_X}$, the equations\footnote{The equations $I^i = 0$ inherit the symmetry $\widetilde{X}_s$ regardless of their regularity since it generates a flow on $\mathcal{E}_X$ and $\widetilde{X}_s(I^i) = r^i I^i$ for some $r^i\in\mathbb{R}$.} $I^i = 0$ inherit the symmetry 
$\widetilde{X}_s = X_s|_{\mathcal{E}_X}$.

\vspace{0.5ex}
\noindent
\textbf{4.} At least one of the following two conditions holds: either each vector field $X_j$ vanishes at some point\footnote{This is equivalent to the condition that $\mathcal{E}_X$ admits an $X_j$-invariant solution.} of $\mathcal{E}_X\subset \mathcal{E}$ for $j = 1, \ldots, k$, or each $dI^i$ vanishes at some point of $\mathcal{E}_X$ for $i = 1, \ldots, l$.

\vspace{0.5ex}
\noindent
In this case, the equations $I^i = 0$ also inherit the symmetries $\widetilde{X}_j = X_j|_{\mathcal{E}_X}$ for $j = 1, \ldots, k$. Indeed, by Theorem~\ref{Theorem1}, the functions $\widetilde{X}_j(I^i)$ are constants and hence vanish.

Consider the submanifold $\mathcal{S}\subset \mathcal{E}_X$ defined by the conditions
\begin{align}
\mathcal{S}\,\colon\qquad
I^1 = \ldots = I^l = 0\,,\qquad dI^1\wedge\ldots\wedge dI^l\neq 0\,,\qquad \widetilde{X}_1 \wedge \ldots \wedge \widetilde{X}_k\neq 0\,.
\label{Sdef}
\end{align}
Finally, we also assume that

\vspace{0.5ex}
\noindent
\textbf{5.} $\dim \mathcal{S} = k+n$ (i.e., that $\mathcal{S}$ is not empty).

\vspace{0.5ex}
\noindent
Then $\mathcal{S}$ inherits the Cartan distribution\footnote{Thus $\mathcal{S}$ can be interpreted as a differential equation on its own. By its solutions we mean (connected) maximal integral manifolds.} from $\mathcal{E}_X$. Since the vector fields $\widetilde{X}_s$, $\widetilde{X}_1$, $\ldots$, $\widetilde{X}_k$ are tangent to $\mathcal{S}$, they determine the respective vector fields $Y_s$, $Y_1$, $\ldots$, $Y_k$ on $\mathcal{S}$. These vector fields preserve the distribution on $\mathcal{S}$, and the following straightforward observation holds (for consistency, we provide its proof in Appendix~\ref{App:A}).

\vspace{1ex}

\lemmaa{\label{Lemma1} There exist functions $f^i\in C^{\infty}(\mathcal{S})$ such that $Y_s = f^i Y_i$. These functions are constant on solutions of $\mathcal{S}$. The differentials $df^1, \ldots, df^k$ are linearly independent at every point of $\mathcal{S}$.}

\vspace{1ex}
\noindent
Thus every solution of $\mathcal{S}$ is invariant under a symmetry of the form $Y_s - C^i Y_i$ for some $C^1$, $\ldots$, $C^k\in \mathbb{R}$. The corresponding solutions of $\mathcal{E}$ are invariant under the respective two-dimensional algebras spanned by $X$ and $X_s - C^i X_i$. In particular, they satisfy the corresponding equations
\begin{align*}
\varphi_s - C^i \varphi_i = 0\,.
\end{align*}
Let us note that all solutions of $\mathcal{S}$ are embedded submanifolds; the general solution of $\mathcal{S}$ is given by the system
\begin{align*}
f^1 = C^1,\quad \ldots,\quad f^k = C^k.
\end{align*}
The corresponding solutions of $\mathcal{E}_X$ are also maximal integral manifolds since the (local) flow of the lift of any vector field from $M$ to $\mathcal{E}_X$ preserves all conditions in~\eqref{Sdef}. 
The image of the restriction of $\pi_{\mathcal{E}}$ to such a solution of $\mathcal{E}_X\subset \mathcal{E}$ is an open subset $N\subset M$. If this restriction is injective, the solution of $\mathcal{E}_X$ defines a unique solution of $\pi_{\mathcal{E}}$ on $N$ (a local one if $N\neq M$, but its domain cannot be extended).

\vspace{1ex}

\remarka{This approach extends to the case where the linear operator defined by $\mathcal{L}_{X_s}$ on the vector space spanned by the cohomology classes of $\omega^1$, $\ldots$, $\omega^l$ is not necessarily diagonalizable, provided that its sum with the operator of multiplication by $a$ is non-degenerate.~\label{Remongeneralization}}

\vspace{1ex}

\remarka{For any $C^1, \ldots, C^k\in \mathbb{R}$, all solutions of $\mathcal{E}$ invariant under both $X$ and $X_s - C^i X_i$ lie on the common level set $I^1 = \ldots = I^l = 0$. This follows from the fact that $\widetilde{X}_s - C^i \widetilde{X}_i$ acts on each $I^j$ as multiplication by a nonzero real number, while it vanishes on such solutions. This result remains valid regardless of the dimension of $\mathcal{E}_X$. \label{Rem_allinvsolareontheleaf}
}

\vspace{1ex}

Symmetry reductions of differential equations can inherit presymplectic structures and Poisson brackets (see~\cite{druzhkovcheviakov2025invariantreductionpartialdifferential, druzhkov2025invariantreductionpartialdifferential}). This observation can be used to select suitable symmetries and conservation laws that determine nonempty systems serving as $\mathcal{S}$ in~\eqref{Sdef}. In some situations, it can also simplify an analysis of the inequalities in~\eqref{Sdef}. In particular, it may suffice to impose only one of them (see Section~\ref{SectionPotenBouss}). 

\section{Examples \label{Sectionexamples}}

Let us illustrate the results of Section~\ref{Mainsection} and Section~\ref{SectionScalsymexectsols}.

In Section~\ref{LRTsubsec}, we consider one example where the emergence of invariance of a conservation law allows us to obtain exact solutions. Notably, the conservation law belongs to a family parametrized by an arbitrary function of one independent variable. We show that such conservation laws, alongside others, can be effectively used for constructing invariant solutions, even if they lack invariance with respect to the whole symmetry subalgebra.

Section~\ref{SectionPotenBouss} demonstrates the approach in Section~\ref{SectionScalsymexectsols} employing the Poisson bivector arising as the reduction of a local Hamiltonian operator of the potential Boussinesq system.

\subsection{The equation of potential nonstationary transonic gas flows \label{LRTsubsec}}

Let us denote by $\mathcal{E}$ the infinite prolongation of the Lin--Reissner--Tsien equation~\cite{lin1948two}.
\begin{align}
w_{yy} - 2w_{tx} - w_x w_{xx} = 0\,.
\label{LRT}
\end{align}
Here $w$ denotes the dependent variable, $x^1 = t$, $x^2 = x$, $x^3 = y$, $F = w_{yy} - 2w_{tx} - w_x w_{xx}$. As coordinates on $\mathcal{E}$, we take all the variables except $w_{yy}$ and its total derivatives.

Consider the following point symmetries
\begin{align*}
&W_0 = -2t\partial_t - y\partial_y + 2w\partial_w\,,\qquad W_{\gamma} = \gamma\partial_x + 2(\dot{\gamma}x + \ddot{\gamma}y^2)\partial_w\,,\qquad W_g = g\partial_w\,,\\
&W_1 = t^3\partial_t + (t^2 x + 2t y^2)\partial_x + 2t^2 y\partial_y + (4x y^2 + 2t x^2 - t^2 w)\partial_w\,,\\
&W = 2ty\partial_x + t^2\partial_y + 4xy\partial_w\,.
\end{align*}
Here $g = g(t)$ is an arbitrary smooth function, $\gamma = \gamma_1 t + \gamma_2 t^{-1}$, $\gamma_1, \gamma_2 \in \mathbb{R}$. We assume that $t\neq 0$.
A complete classification of point symmetries of~\eqref{LRT} can be found in~\cite{Mamontov, ovsiannikov1982group, Ibragimov2023CRCHO}.

The following commutator relations hold:
\begin{align*}
[W_0, W] = -3W,\quad [W_{\gamma}, W] = 0\,,\quad [W_{g}, W] = 0\,,\quad [W_1, W] = 0\,.
\end{align*}
We take $X = E_{\varphi}|_{\mathcal{E}}$ to be the evolutionary symmetry corresponding to $W$. Here
\begin{align*}
&\varphi = 4xy - 2ty w_x - t^2 w_y\,.
\end{align*}
As $X_s = E_{\varphi_s}|_{\mathcal{E}}$, we take the evolutionary symmetry corresponding to $W_0 + W_{\gamma} + W_g + A W_1$, where $A\in \mathbb{R}$. Then $[X_s, X] = aX$ for $a = -3$ and
\begin{align*}
\varphi_s =\ &2w + 2tw_t + yw_y + 2(\dot{\gamma}x + \ddot{\gamma}y^2) - \gamma w_x + g\\
&+ A(4x y^2 + 2t x^2 - t^2 w - t^3 w_t - (t^2 x + 2t y^2) w_x - 2t^2 y w_y)\,.
\end{align*}

\remarka{Solutions of the equation~\eqref{LRT} invariant under both scaling symmetries $W_0$ and $3t\partial_t + x\partial_x + 2y\partial_y - w\partial_w$ are considered in~\cite{haussermann2011self}.}

\subsubsection{The emergence of invariance of a conservation law reduction \label{SubecLRTemerginv}}


The system $\mathcal{E}$ admits the conservation law represented by the following two-component form
\begin{equation}\label{eq:CL:LRT}
\omega = \dfrac{w_y}{t} dt\wedge dx + \dfrac{1}{t}\Big(\dfrac{w_x^2}{2} + 2w_t\Big) dt\wedge dy\,.
\end{equation}
It is associated with the cosymmetry $\,\overline{\!\psi} = {1}/{t}$.
Using it, one can show (see~\cite{VinKr}) that the conservation law is $X$-invariant, while the Lie derivative $\mathcal{L}_{X_s}$ acts on it as multiplication by $b = 3$. For instance, one has $E_{\varphi}(F) = \Phi(F)$ for the operator $\Phi = -t^2 D_y$.
In terms of cosymmetries, the action of $\mathcal{L}_{X}$ on conservation laws corresponds to the action of the operator $X + \Phi^*|_{\mathcal{E}}$. One can see that $(X + \Phi^*|_{\mathcal{E}})\,\overline{\!\psi} = 0$ and hence the conservation law is $X$-invariant. Similarly, the operator $X_s + \Phi_s^*|_{\mathcal{E}}$ corresponding to $\mathcal{L}_{X_s}$, $E_{\varphi_s}(F) = \Phi_s(F)$, multiplies $\,\overline{\!\psi}$ by $b = 3$.

\vspace{1ex}

\remarka{The conservation law \eqref{eq:CL:LRT} belongs to the family of conservation laws associated with cosymmetries of the form $\,\overline{\!\psi} = h(t)$, where $h$ is an arbitrary smooth function.}

\vspace{1ex}

The following one-component form is a solution to the equation $\mathcal{L}_X \omega = d_0 \vartheta$.
\begin{align*}
\vartheta = \dfrac{1}{t} \Big(2tw + 2tyw_y + t^2 \Big(\dfrac{w_x^2}{2} + 2w_t\Big) - 2x^2\Big) dt\,.
\end{align*}
The system $\mathcal{E}_X$ is the infinite prolongation of the equations
\begin{align*}
w_y = \dfrac{4xy - 2ty w_x}{t^2}\,,\qquad w_{tx} = -\dfrac{1}{2}w_x w_{xx} + \dfrac{2t y^2 w_{xx} - t^2 w_x + 2tx - 4y^2}{t^3}\,.
\end{align*}
The restriction $\vartheta|_{\mathcal{E}_X}$ represents the reduction of the conservation law. Since $a + b = 0$, according to Theorem~\ref{Theorem1}, this reduction is $\widetilde{X}_s$-invariant, where $\widetilde{X}_s = X_s|_{\mathcal{E}_X}$, although we do not address the question of its nontriviality. Its study can be based on a compatibility complex for the linearization operator $l_{\mathcal{E}_X}$ (see~\cite[Corollary 7.4]{krasil1998homological}) or on the variation of this element of $E^{\hspace{0.1ex} 0, \hspace{0.15ex} 1}_1(\mathcal{E}_X)$ within an appropriate class of $1$-dimensional submanifolds of $\mathcal{E}_X$. 

\subsubsection{Exact solutions invariant under both $X$ and $X_s$ \label{SubsecLRTsolutions}}

The equation $\mathcal{L}_{\widetilde{X}_s}(\vartheta|_{\mathcal{E}_X}) = d_0 \mu$ admits the following solution
\begin{align*}
&\mu = (A t^2 - 2)\Big(4y^2 w_x - \dfrac{t^2 w_x^2}{2} - 2t (tw_t + w) - \dfrac{8xy^2}{t} + 2x^2\Big) + 2t g\,.
\end{align*}
Thus $\mu$ is constant on solutions of $\mathcal{E}$ invariant under both $X$ and $X_s$.

Since the symmetry\footnote{The equation~\eqref{LRT} is an Euler--Lagrange equation. The symmetry $\frac{1}{t}\partial_w$ corresponds to the conservation law represented by $\omega$.} $\frac{1}{t}\partial_w$ commutes with $X$ and $X_s$, the system $\mathcal{E}_{X, X_s}$ describing solutions invariant under both $X$ and $X_s$ inherits it. Let us introduce the following dependent variable $\eta = tw$. Then
\begin{align*}
&\mu = (A t^2 - 2)\Big(\dfrac{4y^2}{t} \eta_x - \dfrac{\eta_x^2}{2} - 2t \eta_t - \dfrac{8xy^2}{t} + 2x^2\Big) + 2t g\,.
\end{align*}
Assuming that $A t^2 - 2\neq 0$ (e.g., that $A\leqslant 0$), we can write the family of systems $\varphi_s = 0$, $\varphi = 0$, $\mu = C\in \mathbb{R}$, having the same set of solutions as $\mathcal{E}_{X, X_s}$, in the form
\begin{align*}
&(At^2 - 2) \eta_t = \Big(A(2y^2 - tx) - \dfrac{2y^2}{t^2} - \dfrac{\gamma}{t} \Big) \eta_x + 2(\dot{\gamma}x + \ddot{\gamma}y^2) + g + \dfrac{4xy^2}{t^2} + A(2t x^2 - 4x y^2)\,,\\
&\eta_y = \dfrac{4xy - 2y \eta_x}{t}\,,\qquad h_2 \eta_x^2 + h_1 \eta_x + h_0 = 0\,,
\end{align*}
where
\begin{align*}
&h_2 = -\dfrac{A t^2 - 2}{2}\,,\ \ h_1 = 2At^2x - \dfrac{4y^2}{t} + 2\gamma\,,\ \ h_0 = - 4t(\dot{\gamma}x + \ddot{\gamma}y^2) + \dfrac{8xy^2}{t} - 2At^2x^2 - 4x^2 - C\,.
\end{align*}
The corresponding local solutions exist on an open simply connected subset\footnote{If the $1$-form~\eqref{xiform} is exact on a connected component of the open subset~\eqref{Velodomain} for some values of $A$, $\gamma_1$, $\gamma_2$, $C$, then the corresponding solutions exist on it.} $N$ of
\begin{align}
\{(t,x,y)\in\mathbb{R}^3\,\colon\ t\neq 0\,,\ At^2 - 2 \neq 0\,,\ h_1^2 - 4h_2h_0 > 0 \}\,.
\label{Velodomain}
\end{align}
Here, $h_1^2 - 4h_2h_0$ can be written in the form
\begin{align*}
h_1^2 - 4h_2h_0 = \Big(4x - \dfrac{4y^2}{t} + 2\gamma_1 t - \dfrac{2\gamma_2}{t} + 2A\gamma_2t\Big)^2 + (4 - 2At^2)(C + 2A\gamma_2^2 + 4\gamma_1\gamma_2)\,.
\end{align*}
The solutions are given by the formula
\begin{align*}
w = \dfrac{1}{t}\int_{\Gamma} \xi\,,
\end{align*}
where $\Gamma$ is a smooth path in $N$ connecting a chosen point $(t_0, x_0, y_0)\in N$ and $(t, x, y)\in N$,
\begin{align}
\begin{aligned}
\xi = &\ \dfrac{1}{At^2 - 2}\bigg(\dfrac{-h_1\pm \sqrt{h_1^2 - 4h_2h_0}}{2h_2} \Big(A(2y^2 - tx) - \dfrac{2y^2}{t^2} - \dfrac{\gamma}{t} \Big) + 2(\dot{\gamma}x + \ddot{\gamma}y^2) + g + \dfrac{4xy^2}{t^2}\\
&+ A(2t x^2 - 4x y^2) \bigg)dt + \dfrac{-h_1\pm \sqrt{h_1^2 - 4h_2h_0}}{2h_2}\, dx - \dfrac{2y}{t}\Big(\dfrac{-h_1\pm \sqrt{h_1^2 - 4h_2h_0}}{2h_2} - 2x\Big)dy\,.
\end{aligned}
\label{xiform}
\end{align}

\begin{figure}[h!]
\includegraphics[width=0.5\textwidth]{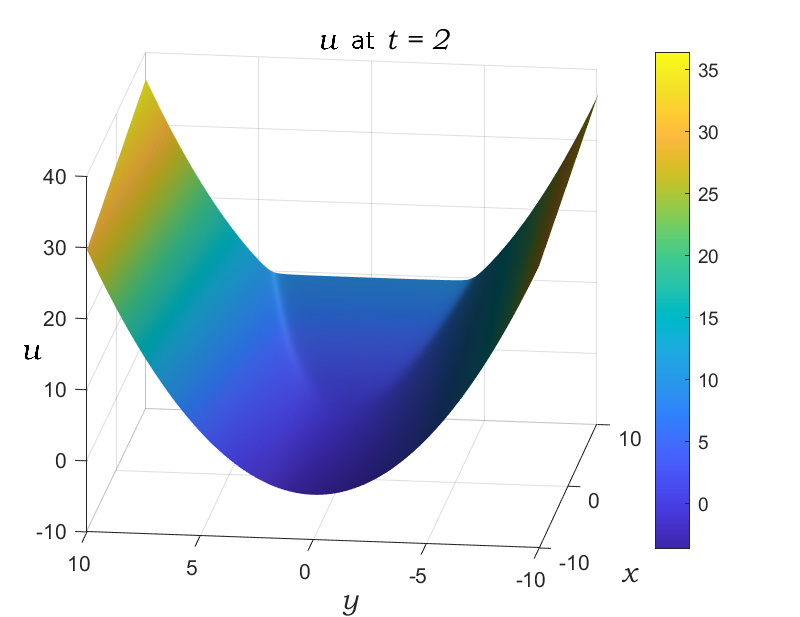}
\hfill
\raisebox{0pt}
{\includegraphics[width=0.5\textwidth]{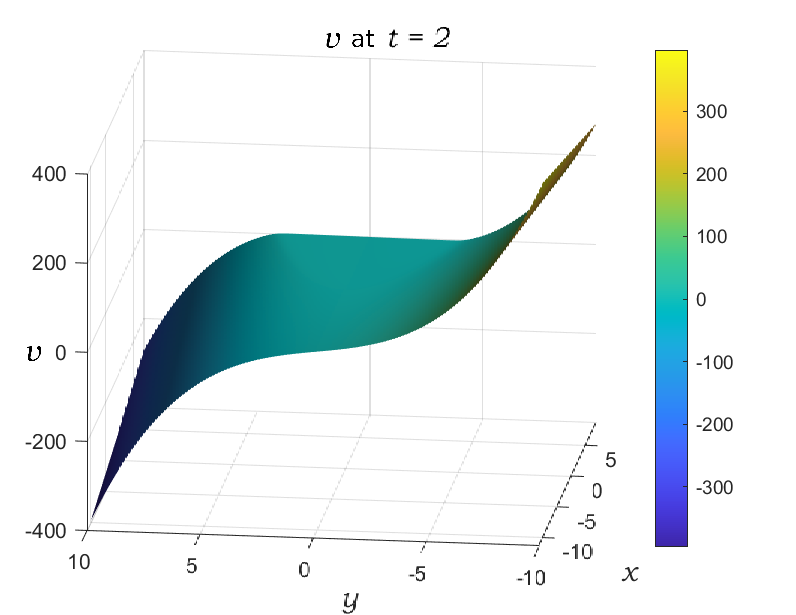}}
\caption{$u$ and $v$ at $t = 2$ for $A = -1$, $\gamma_1 = \gamma_2 = 1$, $C = 0$.}
\label{fig1}
\end{figure}

Let us note that from a physical point of view, $w$ plays the role of the potential~\cite{lin1948two}. The corresponding velocity $\mathbf{v} = (u, v)$ is defined on~\eqref{Velodomain}. Its components $u = w_x$ and $v = w_y$ are given by the formulas
\begin{align*}
u = \dfrac{-h_1\pm \sqrt{h_1^2 - 4h_2h_0}}{2h_2t}\,,\qquad v = \dfrac{4xy}{t^2} - y \dfrac{-h_1\pm \sqrt{h_1^2 - 4h_2h_0}}{h_2t^2}\,.
\end{align*}
On the domain, $u(t, x, y) = u(t, x, -y)$ and $v(t, x, y) = -v(t, x, -y)$. For example, the values $A = -1$, $\gamma_1 = \gamma_2 = 1$, $C = 0$ correspond to $u$ and $v$ defined on $\{(t, x, y)\in \mathbb{R}^3\,\colon\ t\neq 0\}$. Their graphs (at $t = 2$) are presented in Figure~\ref{fig1}. At each point $(x, y)$, both components of the velocity $\mathbf{v}$ tend to zero as $t\to \pm\infty$.

\noindent

\subsubsection{Numerical validation and stability \label{SectionNumerical}}


We now validate the exact solution numerically by evolving the first-order
system
\begin{equation}\label{eq:system}
  u_t = \tfrac{1}{2}\,v_y - D_x\bigl(\tfrac{1}{4}\,u^2\bigr)\,,
  \qquad v_x = u_y\,,
\end{equation}
satisfied by the solution derivatives $u = w_x$, $v = w_y$ of the PDE \eqref{LRT}.
The exact solution belongs to the family parameterised by
$(A,\gamma_1,\gamma_2,C)=(-1,1,1,0)$ and is given in closed form;
it is smooth for all $t>0$ on the computational domain
$\Omega=[-10,10]^2$. Initial data at $t_0=2$ and Dirichlet boundary conditions on
$\partial\Omega$ are prescribed from the exact solution.

The evolution system~\eqref{eq:system} contains a nonlinear hyperbolic
flux $f(u)=u^2/4$.  We discretise this term with the
fifth-order weighted essentially non-oscillatory (WENO5) finite-difference
reconstruction of Jiang and Shu~\cite{JiangShu1996}, combined with the
Rusanov (local Lax--Friedrichs) numerical flux~\cite{Toro2009} to
provide upwind dissipation at each cell interface.  The characteristic
speed is $f'(u)=u/2$, and the local maximum wave speed
determines both the numerical viscosity coefficient in the Rusanov flux
and the adaptive CFL time-step restriction. All spatial derivatives
in~$y$ are computed with standard
fourth-order centred finite differences in the interior, with
second-order one-sided stencils at the two boundary-adjacent grid
lines~\cite{LeVeque2007}. The variable~$v$ is not evolved independently.  Instead, it is
reconstructed at each step from the integrability
constraint $v_x = u_y$ by setting $v(t,x_{\min},y)$ to its exact value
and integrating
\[
v(t,x,y) = v(t,x_{\min},y) + \int_{x_{\min}}^{x} u_y(t,\xi,y)\,
d\xi
\]
across the domain using composite Simpson quadrature.  This projection
step enforces the constraint up to the quadrature
and differentiation truncation errors, and avoids the need for a
separate evolution equation or Lagrange-multiplier coupling for~$v$.

Time integration is performed with the three-stage, third-order strong
stability-preserving Runge--Kutta method SSPRK(3,3) of Shu and
Osher~\cite{ShuOsher1988}, which preserves the total-variation-diminishing
property of the underlying spatial discretization under the CFL
restriction $\Delta t \le \Delta t_{\mathrm{FE}}$, where
$\Delta t_{\mathrm{FE}}$ is the forward-Euler stability limit.  We use the spatial domain $\Omega = [-10,10]^2$ with a $301\times 301$ uniform grid and a
CFL number $\nu = 0.25$ throughout, yielding an adaptive time step that
grows from $\Delta t \approx 1.2\times10^{-3}$ near $t_0=2$ to
$\Delta t \approx 10^{-2}$ by $t=6$ as the solution amplitude decays
(see Table~\ref{tab:weno}).  A comprehensive account of SSP methods
and their optimality properties can be found in the monograph by
Gottlieb, Ketcheson, and Shu~\cite{GottliebKetchesonShu2011}.

\begin{table}[ht]
\centering
\caption{Diagnostic output of the WENO5--SSPRK(3,3) scheme. The solution is initialized at $t_0 = 2$
from the exact solution and integrated to $t_f = 8$. Listed are the time step index, current time, adaptive time step, the relative $L^2$ and $L^\infty$ errors of $u$, and the constraint residual $R_{\mathrm{curl}} = \|u_y - v_x\|_{L^\infty(\Omega_{\mathrm{int}})}$ on the interior grid.}
\label{tab:weno}
\begin{tabular}{c c c c c c}
\hline
Step & $t$ & $\Delta t$ & $E^u_{2,\mathrm{rel}}$ & $E^u_{\infty,\mathrm{rel}}$ & $R_{\mathrm{curl}}$ \\
\hline
200  & 2.211 & $1.22 \times 10^{-3}$ & $2.41 \times 10^{-4}$ & $1.15 \times 10^{-3}$ & $9.13 \times 10^{-3}$ \\
400  & 2.501 & $1.71 \times 10^{-3}$ & $4.48 \times 10^{-4}$ & $1.42 \times 10^{-3}$ & $7.65 \times 10^{-3}$ \\
600  & 2.921 & $2.56 \times 10^{-3}$ & $7.41 \times 10^{-4}$ & $1.88 \times 10^{-3}$ & $8.13 \times 10^{-3}$ \\
800  & 3.569 & $4.04 \times 10^{-3}$ & $1.02 \times 10^{-3}$ & $3.07 \times 10^{-3}$ & $1.37 \times 10^{-2}$ \\
1000 & 4.602 & $6.45 \times 10^{-3}$ & $1.04 \times 10^{-3}$ & $4.17 \times 10^{-3}$ & $8.72 \times 10^{-3}$ \\
1200 & 6.218 & $9.85 \times 10^{-3}$ & $9.15 \times 10^{-4}$ & $4.70 \times 10^{-3}$ & $7.75 \times 10^{-3}$ \\
1356 & 8.000 & $5.25 \times 10^{-3}$ & $7.66 \times 10^{-4}$ & $7.05 \times 10^{-3}$ & $5.31 \times 10^{-3}$ \\
\hline
\end{tabular}
\end{table}

The results are summarized in Table~\ref{tab:weno}. Over the integration window $t \in [2, 8]$, the relative $L^2$ error $E^u_{2,\mathrm{rel}} = \|u - u_{\mathrm{ex}}\|_{L^2(\Omega)}/\|u_{\mathrm{ex}}\|_{L^2(\Omega)}$ remains at the level of $O(10^{-4})$ to $O(10^{-3})$. The relative pointwise error $E^u_{\infty,\mathrm{rel}}$ grows from $1.15 \times 10^{-3}$ to $7.05 \times 10^{-3}$ over the integration, remaining below $1\%$ throughout. Crucially, neither error norm exhibits any exponential or algebraic blow-up, confirming that the exact solution is dynamically stable and can be tracked by standard high-resolution schemes over extended time horizons. The constraint residual $R_{\mathrm{curl}} = \|u_y - v_x\|_{L^\infty(\Omega_{\mathrm{int}})}$ stays at the level of $O(10^{-3})$ to $O(10^{-2})$, consistent with the truncation error of the finite-difference and quadrature operators, and shows no secular growth. The total integration requires 1356 time steps and completes in approximately 22 seconds on a single core, demonstrating that the combination of WENO5 spatial discretization with SSPRK(3,3) time integration provides an efficient and robust computational framework.

\begin{figure}[t]
\centering
\includegraphics[width=\textwidth]{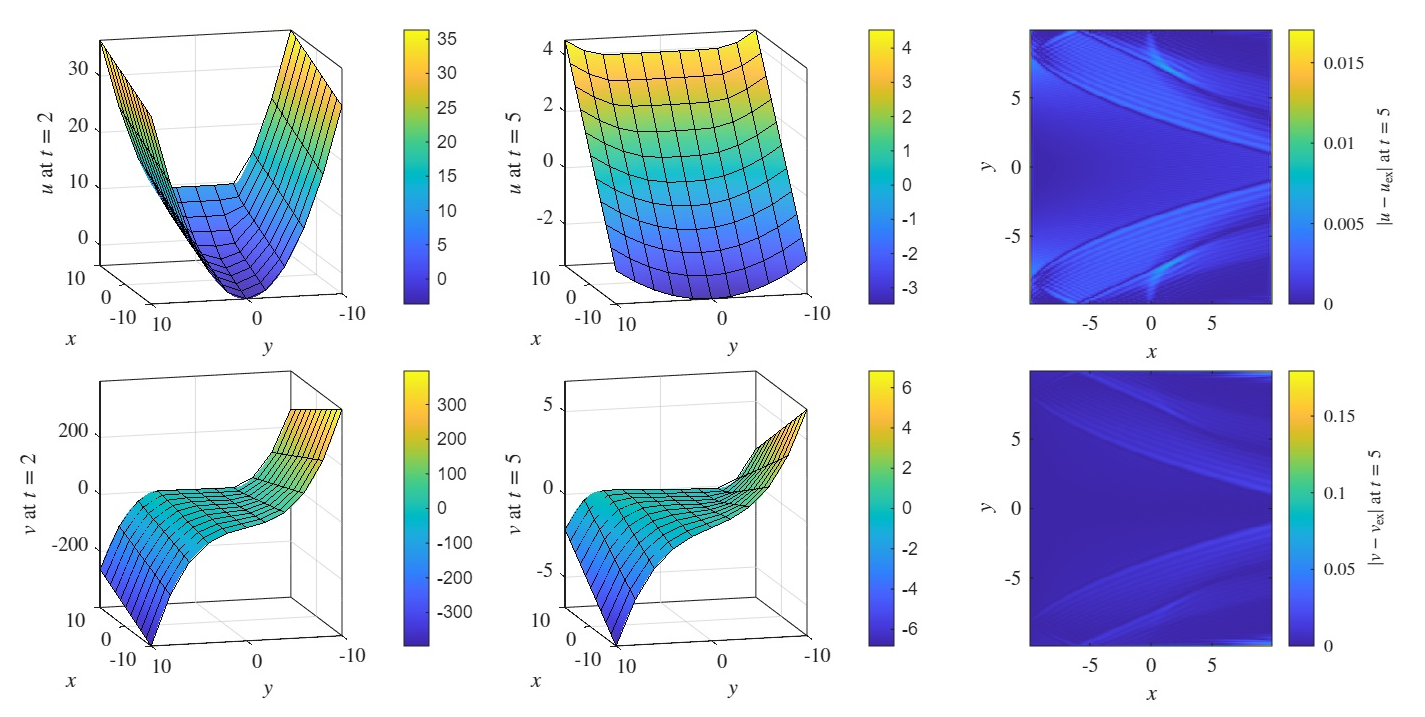}
\caption{Exact solution $u(t,x,y)$ and $v(t,x,y)$ at $t=2$ (first column) and $t=5$ (second column). The third column displays the absolute interior errors $|u-u_{\mathrm{ex}}|$ and $|v-v_{\mathrm{ex}}|$ at $t=5$ over the computational domain excluding two boundary layers.}
\label{fig:uv_surfaces_errors}
\end{figure}

Figure~\ref{fig:uv_surfaces_errors} presents the exact solution profiles for $u$ and $v$ at two representative times, $t=2$ and $t=5$. The surfaces illustrate the smooth spatial structure and its temporal evolution. The rightmost panels show the absolute numerical errors at $t=5$ at interior grid points. The error magnitude remains small relative to the solution amplitude, confirming stability and accuracy of the discretization at this resolution.

\begin{figure}[htbp]
\centering
\includegraphics[width=0.75\textwidth]{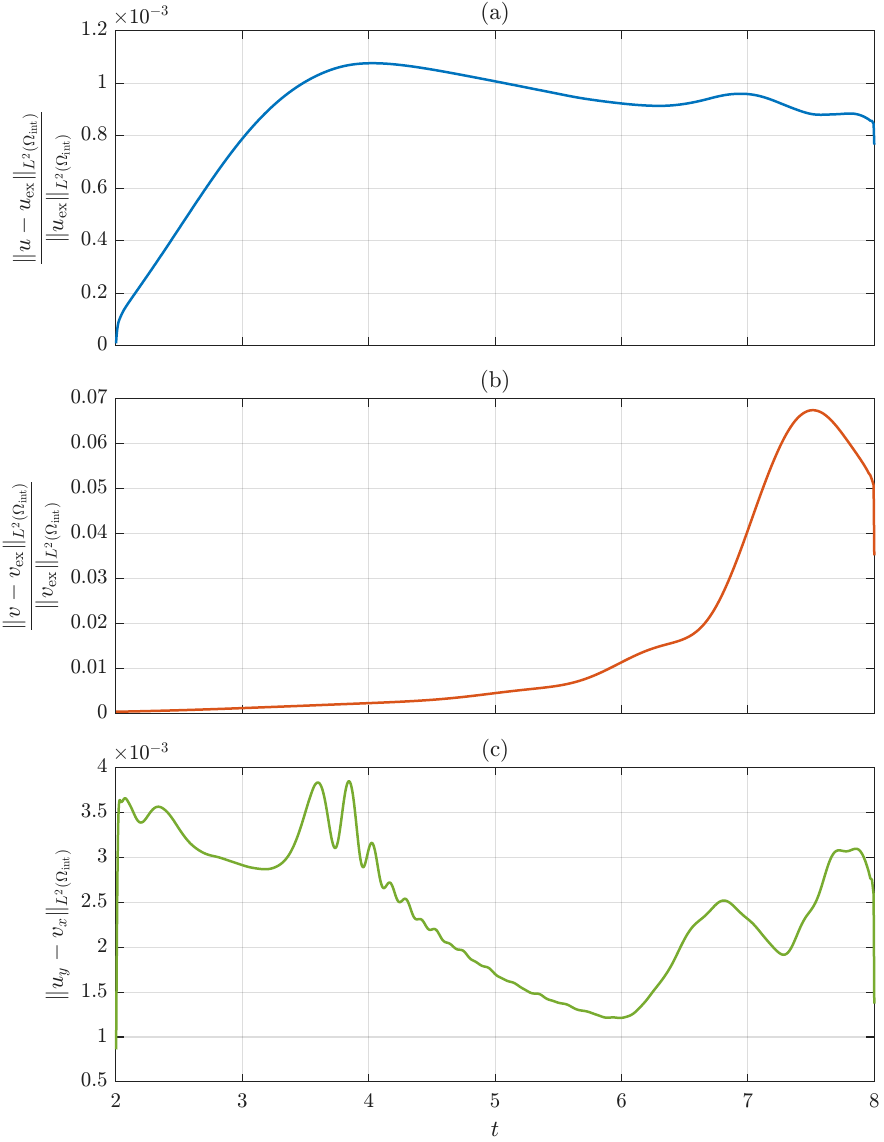}
\caption{Time history of interior $L^2$ error metrics on
$\Omega_{\mathrm{int}}$ (the interior grid with two boundary layers removed on
each side).
(a)~Relative error of~$u$.
(b)~Relative error of the reconstructed~$v$.
(c)~Constraint residual $\|u_y - v_x\|_{L^2(\Omega_{\mathrm{int}})}$.
Parameters: $301\times 301$ grid, $\mathrm{CFL}=0.25$, $t\in[2,8]$.}
\label{fig:L2errors}
\end{figure}

Figure~\ref{fig:L2errors} shows the full time history of the relative $L^2$ errors, evaluated on the interior
subdomain $\Omega_{\mathrm{int}}$ (two grid layers removed from each
boundary, so that all centred stencils are fully supported).
Panel~(a) confirms the gradual, sub-linear growth of the $u$-error
already seen in Table~\ref{tab:weno}. Panel~(b) shows that the reconstructed field~$v$, obtained solely from the constraint projection, tracks its exact counterpart with comparable
relative accuracy despite accumulating both differentiation and
quadrature errors. Panel~(c) displays the $L^2$ constraint residual
$\|u_y-v_x\|_{L^2(\Omega_{\mathrm{int}})}$, which plateaus at a level
set by the spatial truncation error and exhibits no secular drift,
indicating that the projection step maintains consistency with the
integrability condition throughout the integration.

\subsection{The potential Boussinesq system \label{SectionPotenBouss}}


Let $\mathcal{E}$ denote the infinite prolongation of the potential Boussinesq system
\begin{align*}
u_t = v_x\,,\qquad v_t = \dfrac{8}{3}uu_x + \dfrac{1}{3}u_{xxx}\,.
\end{align*}
Here $u^1 = u$, $u^2 = v$, $x^1 = t$, $x^2 = x$, $F^1 = u_t - v_x$, $F^2 = v_t - \frac{8}{3}uu_x - \frac{1}{3}u_{xxx}$. We treat $t$, $x$, $u$, $v$, and all total $x$-derivatives $u_x$, $v_x$, $\ldots$ as coordinates on $\mathcal{E}$. This gives rise to the corresponding inclusion $\mathcal{F}(\mathcal{E})\subset \mathcal{F}(\pi)$ and allows us to identify the operator $D_x|_{\mathcal{E}}$ with $D_x$.

This system admits the point symmetry
\begin{align*}
Z_s  = -2t\partial_t - x\partial_x + 2u\partial_u + 3v\partial_v\,
\end{align*}
corresponding to scaling transformations. Let $X_s$ denote the corresponding evolutionary symmetry. The components of its characteristic are given by
\begin{align*}
\varphi_s^1 = 2u + x u_x + 2t v_x\,,\qquad \varphi_s^2 = 3v + x v_x + \dfrac{2t}{3} (8uu_x + u_{xxx})\,.
\end{align*}

The potential Boussinesq system admits a local Hamiltonian operator
\begin{align*}
\nabla =
\begin{pmatrix}
0 & D_x\\
D_x & 0
\end{pmatrix}
\end{align*}
The cosymmetry $\psi = (\psi_1, \psi_2)$ given by
\begin{align*}
\psi_1 = \dfrac{1}{3}u_{4x} + 4uu_{xx} + 2u_x^2 + \dfrac{32}{9} u^3 + 2v^2,\qquad
\psi_2 = v_{xx} + 4uv
\end{align*}
corresponds to a conservation law. Here we adopt the notation $u_{4x} = u_{xxxx}$, $u_{5x} = u_{xxxxx}$, $\ldots$ As $X$, we take the corresponding Hamiltonian symmetry. Then $[X_s, X] = aX$ for $a = 4$. Here $X = E_{\varphi}|_{\mathcal{E}}$ for $\varphi = \nabla(\psi) = (\varphi^1, \varphi^2)$,
\begin{align*}
\varphi^1 = D_x(\psi_2)\,,\qquad \varphi^2 = D_x(\psi_1)\,.
\end{align*}
The following five cosymmetries $\psi^i = (\psi_1^i, \psi_2^i)$ of $\mathcal{E}$ correspond to $X$-invariant conservation laws.
\begin{align*}
&\psi_1^1 = v\,,\ \ \psi_2^1 = u\,,\qquad
\psi_1^2 = \dfrac{4}{3}u^2 + \dfrac{1}{3}u_{xx}\,,\ \ \psi_2^2 = v\,,\\
&\psi_1^3 = \dfrac{1}{3}v_{4x} + \dfrac{10}{3} u v_{xx} + \dfrac{5}{3} u_x v_x + \dfrac{5}{3} u_{xx} v + \dfrac{20}{3} u^2 v\,,\ \
\psi_2^3 = \dfrac{1}{3}u_{4x} + \dfrac{5}{2} u_x^2 + \dfrac{20}{9} u^3 + \dfrac{5}{2} v^2 + \dfrac{10}{3} u u_{xx}\,.\\
&\psi_1^4 = 1\,,\ \ \psi_2^4 = 0\,,\qquad
\psi_1^5 = 0\,,\ \ \psi_2^5 = 1\,.
\end{align*}
The Lie derivative $\mathcal{L}_{X_s}$ acts on them as multiplication by $4$, $5$, $8$, $1$, and $2$, respectively. None of these coefficients coincides with $-a = -4$. Then the reduction of each of these conservation laws is uniquely represented by a function in $\mathcal{F}(\mathcal{E}_X)$ on which $\widetilde{X}_s = X_s|_{\mathcal{E}_X}$ acts by multiplication with a real number.

Let us take as $X_1$, $X_2$, $X_3$ the Hamiltonian symmetries that correspond to $\psi^1$, $\psi^2$, $\psi^3$, respectively. The five conservation laws are $\mathfrak{g}$-invariant, where the Lie algebra $\mathfrak{g}$ is spanned by $X$, $X_1$, $X_2$, $X_3$. It is commutative, while $[X_s, X_1] = X_1$, $[X_s, X_2] = 2X_2$, $[X_s, X_3] = 5X_3$. Then $\mathcal{L}_{X_s}$ acts on $\mathfrak{g}/\langle X \rangle$ as an invertible linear operator. The restrictions $\widetilde{X}_i = X_i|_{\mathcal{E}_X}$ are determined by
\begin{align*}
&\varphi_1^1|_{\mathcal{E}_X} = u_x\,,\ \varphi_1^2|_{\mathcal{E}_X} = v_x\,,\qquad
\varphi_2^1|_{\mathcal{E}_X} = v_x\,,\ \varphi_2^2|_{\mathcal{E}_X} = \dfrac{8}{3}uu_x + \dfrac{1}{3}u_{xxx}\,,\\
&\varphi_3^1|_{\mathcal{E}_X} = -4u^2 u_x + \dfrac{1}{3}u_x u_{xx} - \dfrac{2}{3}u u_{xxx} + v v_x,\ \varphi_3^2|_{\mathcal{E}_X} = u_x v_{xx} + \dfrac{16}{3} u u_x v - \dfrac{2}{3} u_{xx} v_x + \dfrac{1}{3} u_{xxx} v - \dfrac{4}{3} u^2 v_x.
\end{align*}
As coordinates on $\mathcal{E}_X$, we take $t$, $x$, $u$, $v$, $u_x$, $v_x$, $u_{xx}$, $v_{xx}$, $u_{xxx}$, $u_{4x}$. Each vector field $\widetilde{X}_j$ vanishes at some point. For example, one can take $t = x = u = v = u_x = v_x = u_{xx} = v_{xx} = u_{xxx} = u_{4x} = 0$ for any of them.

The unique functions representing the reductions of the five conservation laws and rescaled by $\widetilde{X}_s$ are given by
\begin{align*}
I^1 =\, &-\dfrac{1}{2}v_x^2 + \dfrac{1}{6}u_{xx}^2 + \dfrac{8}{3}u^4 - \dfrac{1}{3}u_x u_{xxx} + vv_{xx} + \dfrac{1}{3}uu_{4x} + 4u^2u_{xx} + 4uv^2,\\
I^2 =\ &\dfrac{4}{3} v^3 + \dfrac{8}{3} u_x^2 v + \dfrac{4}{3} u^2 v_{xx} + \dfrac{1}{3} u_{4x} v - \dfrac{1}{3} u_{xxx} v_x - \dfrac{8}{3} uu_x v_x + \dfrac{1}{3} u_{xx} v_{xx} + \dfrac{16}{3} u^3 v + 4uu_{xx} v\,,\\
I^3 =\ &10 u u_{xx} v^2 + \dfrac{22}{3} u u_x^2 u_{xx} + \dfrac{20}{3} u^2 v v_{xx} + \dfrac{1}{3} u_{xx} v v_{xx} - u_x v_x v_{xx} + \dfrac{4}{3} u^2 u_x u_{xxx} + \dfrac{10}{9} u u_{xx} u_{4x}\\
&-\dfrac{1}{9} u_x u_{xx} u_{xxx} - \dfrac{1}{3} u_{xxx} v v_x + \dfrac{80}{9} u^3 u_x^2 + \dfrac{80}{9} u^4 u_{xx} + 6 u^2 u_{xx}^2 + \dfrac{160}{9} u^3 v^2 + \dfrac{17}{3} u_x^2 v^2 + \dfrac{2}{3} u^2 v_x^2\\
&+ \dfrac{1}{3} u_{xx} v_x^2 + u v_{xx}^2 + \dfrac{1}{9} u u_{xxx}^2 + \dfrac{20}{27} u^3 u_{4x} + \dfrac{5}{6} u_x^2 u_{4x} + \dfrac{5}{6} u_{4x} v^2 - \dfrac{16}{3} u u_x v v_x + \dfrac{320}{81} u^6 + \dfrac{19}{6} u_x^4\\
&+ \dfrac{1}{27} u_{xx}^3 + \dfrac{5}{2} v^4 + \dfrac{1}{18} u_{4x}^2\,,\\
I^4 =\ &v_{xx} + 4uv\,,\\
I^5 =\ &\dfrac{1}{3}u_{4x} + 4uu_{xx} + 2u_x^2 + \dfrac{32}{9} u^3 + 2v^2,
\end{align*}
respectively. Here $I^1$, $I^2$, $I^3$ correspond to the symmetries $\widetilde{X}_1$, $\widetilde{X}_2$, $\widetilde{X}_3$ through the Poisson bivector inherited from $\nabla$ on $\mathcal{E}$ (see~\cite{druzhkov2025invariantreductionpartialdifferential}). This Poisson bivector has the form
\begin{align*}
&3 \partial_u\wedge \partial_{u_{xxx}} + \partial_{v}\wedge \partial_{v_x} - 3\partial_{u_x}\wedge \partial_{u_{xx}} + 36u\partial_{u_x}\wedge \partial_{u_{4x}} + 4u\partial_{v_x}\wedge \partial_{v_{xx}} + 12v \partial_{v_x}\wedge \partial_{u_{4x}}\\
&- 36u \partial_{u_{xx}}\wedge \partial_{u_{xxx}} - 36u_x \partial_{u_{xx}}\wedge \partial_{u_{4x}} - 12 v \partial_{v_{xx}}\wedge \partial_{u_{xxx}} - (336u^2 - 36u_{xx})\partial_{u_{xxx}}\wedge \partial_{u_{4x}}\,.
\end{align*}
It can be readily seen that the kernel of the Poisson bivector is four-dimensional at every point of $\mathcal{E}_X$; it is spanned by $dI^4$, $dI^5$, $dt$, $dx$. Hence, both inequalities in~\eqref{Sdef} are equivalent to $\widetilde{X}_1\wedge \widetilde{X}_2\wedge \widetilde{X}_3\neq 0$ and the subsystem $\mathcal{S}$ is given by
\begin{align*}
I^1 = \ldots = I^5 = 0\,,\qquad \widetilde{X}_1\wedge \widetilde{X}_2\wedge \widetilde{X}_3\neq 0\,.
\end{align*}

Every solution of $\mathcal{S}$ satisfies the condition $\widetilde{X}_s - C^i \widetilde{X}_i = 0$ for some $C^1, C^2, C^3\in \mathbb{R}$. This condition takes the form of the following system of eight equations
\begin{align}
\begin{aligned}
&\widetilde{D}_x^{\hspace{0.15ex} i}\left(\varphi_s^1|_{\mathcal{E}_X}\right) - C^1 \widetilde{D}_x^{\hspace{0.15ex} i}\left(\varphi_1^1|_{\mathcal{E}_X}\right) - C^2 \widetilde{D}_x^{\hspace{0.15ex} i}\left(\varphi_2^1|_{\mathcal{E}_X}\right) - C^3 \widetilde{D}_x^{\hspace{0.15ex} i}\left(\varphi_3^1|_{\mathcal{E}_X}\right) = 0\,,\qquad i = 0, 1, 2, 3, 4\,,\\
&\widetilde{D}_x^{\hspace{0.15ex} j}\left(\varphi_s^2|_{\mathcal{E}_X}\right) - C^1 \widetilde{D}_x^{\hspace{0.15ex} j}\left(\varphi_1^2|_{\mathcal{E}_X}\right) - C^2 \widetilde{D}_x^{\hspace{0.15ex} j}\left(\varphi_2^2|_{\mathcal{E}_X}\right) - C^3 \widetilde{D}_x^{\hspace{0.15ex} j}\left(\varphi_3^2|_{\mathcal{E}_X}\right) = 0\,,\qquad j = 0, 1, 2\,,
\end{aligned}
\label{AdditSymmInvariance}
\end{align}
where $\widetilde{D}_x^{\hspace{0.15ex} 0}$ is the identity operator, $\widetilde{D}_x = D_x|_{\mathcal{E}_X}$,
\begin{align*}
\widetilde{D}_x =\ &\partial_x + u_x \partial_u + v_x \partial_v + u_{xx} \partial_{u_x} + v_{xx} \partial_{v_x} + u_{xxx} \partial_{u_{xx}} - 4(u_x v + u v_x) \partial_{v_{xx}}\\
&+ u_{4x} \partial_{u_{xxx}} - (24u_x u_{xx} + 12uu_{xxx} + 32 u^2 u_x + 12v v_x)\partial_{u_{4x}}.
\end{align*}
At a point of $\mathcal{S}\subset \mathcal{E}_X$, there are three equations among~\eqref{AdditSymmInvariance}, from which one can eliminate $C^1, C^2, C^3$. Then $dI^1, \ldots, dI^5$ and the differentials of the left-hand sides of these three equations, taken at the corresponding values of $C^1$, $C^2$, $C^3$, are linearly independent at the point.

The following point is in $\mathcal{S}$
\begin{align*}
&t = 1,\, x = 1,\, u = \dfrac{35}{726},\, u_x = \dfrac{23545}{11979},\, u_{xx} = -\dfrac{153685}{131769},\, u_{xxx} = -\dfrac{11485580}{4348377},\, v = \dfrac{160}{363},\, v_x = -\dfrac{21200}{11979}.
\end{align*}
The coordinates $v_{xx}$ and $u_{4x}$ are determined from the equations $I^4 = I^5 = 0$. The corresponding solution of $\mathcal{E}_X$ is the connected component of the manifold determined by the system of thirteen algebraic equations $I^1 = \ldots = I^5 = 0$ and~\eqref{AdditSymmInvariance}, subject to $C^1 = C^2 = 0$, $C^3 = 1$ and the inequality $\widetilde{X}_1\wedge \widetilde{X}_2 \wedge \widetilde{X}_3\neq 0$.

\vspace{1ex}

\remarka{One can use the symmetries $\partial_x$ and $\partial_t$ to obtain solutions with nonzero $C^1$ and $C^2$. Similarly, the scaling symmetry can be applied to obtain solutions with other nonzero $C^3$.}



\vspace{1ex}

\remarka{All solutions on the common level set $I^1 = \ldots = I^5 = 0$ that do not satisfy the inequality $\widetilde{X}_1\wedge \widetilde{X}_2 \wedge \widetilde{X}_3\neq 0$ are invariant under some symmetries of the form $A^i \widetilde{X}_i$ for $A^1, A^2, A^3\in \mathbb{R}$. This follows from the observation that the flow of the lift of a vector field from $M$ to $\mathcal{E}_X$ preserves any vector field of the form $A^i \widetilde{X}_i$. 
Thus, all solutions on $I^1 = \ldots = I^5 = 0$ are invariant under some symmetries additional to $X$. \label{Rem_allsolareinvar}}

\vspace{1ex}

We do not address the practical utility of the resulting description of solutions in terms of eight algebraic equations --- $I^1 = \ldots = I^5 = 0$ together with three equations, dynamically chosen from the eight equations~\eqref{AdditSymmInvariance}, depending on the point along the solution. It should be noted, however, that all thirteen equations are satisfied throughout the domains of these solutions and, together, completely determine them. Such descriptions may serve as a tool for validating numerical methods that approximate invariant solutions, for example, via the flows of combinations of the total derivatives $D_{x^i}|_{\mathcal{E}_X}$.


\section{Discussion}

In this paper, we have extended the invariant reduction mechanism developed in Parts~I--III~\cite{druzhkov2025invariant, druzhkovcheviakov2025invariantreductionpartialdifferential, druzhkov2025invariantreductionpartialdifferential} to geometric structures that are not preserved by a symmetry but are \emph{rescaled} by it. The central result is Theorem~\ref{Theorem1}, which shows that when a symmetry $X_s$ satisfies $[X_s,X]=aX$ and acts on an $X$-invariant element of $E^{p,q}_1(\mathcal{E})$ as multiplication by~$b$, the induced action of $\widetilde{X}_s = X_s|_{\mathcal{E}_X}$ on the reduced class shifts the eigenvalue to $a+b$, provided the reduction is well-defined. This shift rule has two immediate structural consequences. The first is the \emph{emergence of invariance}: when $a\neq0$ and $a+b=0$, the reduction acquires an invariance absent at the original level (Remark~\ref{Rem2}), as demonstrated by the conservation law reduction in Section~\ref{SubecLRTemerginv}. The second is the \emph{loss of invariance}: when $a\neq 0$ and $b=0$, an invariant structure fails to remain invariant after reduction if its outcome is nontrivial. Both phenomena are governed by a single arithmetic condition and do not depend on the specific nature of the geometric object being reduced.

Building on the shift rule, we described in Section~\ref{SectionScalsymexectsols} a class of exact solutions to PDE systems that possess sufficiently many symmetries and conservation laws satisfying the compatibility conditions~1--5. These conditions guarantee that the reduced system $\mathcal{E}_X\subset \mathcal{E}$ has a subsystem $\mathcal{S}\subset\mathcal{E}_X$, determined by suitably chosen elements of the group $E^{\hspace{0.1ex} 0,\hspace{0.1ex} 1}_1(\mathcal{E})$, that inherits the scaling symmetry $X_s$. On $\mathcal{S}$, the symmetry $X_s$ decomposes as a linear combination of other symmetries $X_1$, $\ldots$, $X_k$ of $\mathcal{E}$, and the coefficients are functionally independent functions $f^1$, $\ldots$, $f^k$, constant along solutions of $\mathcal{S}$. This decomposition converts the problem of finding suitable $X$-invariant solutions into an explicit algebraic one: the general solution of~$\mathcal{S}$ is given by specifying the values $C^1,\ldots,C^k$ of the functions $f^1,\ldots,f^k$. The resulting solutions of $\mathcal{E}$ are simultaneously invariant under two symmetries, $X$ and $X_s-C^iX_i$, and are completely determined by explicitly constructed functions that are constant on them. Notably, this description is entirely geometric: it relies only on the intrinsic geometry of the PDE system and does not require integrability-related structures such as Lax pairs, B\"acklund transformations, or inverse scattering.

The two examples presented in Section~\ref{Sectionexamples} illustrate complementary aspects of the framework. The Lin--Reissner--Tsien equation~\eqref{LRT} provides a setting in which a conservation law from a family parameterized by an arbitrary function of one independent variable participates in the reduction. The conservation law associated with the cosymmetry $\,\overline{\!\psi} = 1/t$ is $X$-invariant, while the Lie derivative along the symmetry $X_s$ (with $[X_s, X] = -3X$) acts on it as multiplication by $b=3$. Since $a=-3$, one has $a+b=0$, and the reduction acquires scaling invariance despite the conservation law itself not being $X_s$-invariant (Section~\ref{SubecLRTemerginv}). The exact solutions (Section~\ref{SubsecLRTsolutions}) are given by quadrature via the integral of an explicitly constructed closed $1$-form~\eqref{xiform} and are defined on the domains determined by~\eqref{Velodomain} and the parameters $A$, $\gamma_1$, $\gamma_2$, and $C$. The numerical validation in Section~\ref{SectionNumerical} confirms the dynamical stability of a representative solution over a finite time integration window using a WENO5--SSPRK(3,3) scheme, with both the $L^2$ and $L^{\infty}$ errors remaining bounded and the integrability constraint residual $R_{\mathrm{curl}}$ exhibiting no secular growth.

The potential Boussinesq system (Section~\ref{SectionPotenBouss}) illustrates the full machinery of Section~\ref{SectionScalsymexectsols}, including the use of a Poisson bivector inherited through the mechanism of Part~III~\cite{druzhkov2025invariantreductionpartialdifferential}. The Lie derivative $\mathcal{L}_{X_s}$ acts on the five conservation laws with eigenvalues $4$, $5$, $8$, $1$, $2$, none of which coincides with $-a=-4$.
This ensures, by Theorem~\ref{Theorem1}, that their reductions can be represented by rescaled functions $I^1$, $\ldots$, $I^5$ so that the common level set $I^1=\ldots=I^5=0$ inherits the scaling symmetry $\widetilde{X}_s$. The four-dimensional kernel of the Poisson bivector identifies both inequalities in~\eqref{Sdef}, reducing the description of $\mathcal{S}$ to the relations $I^1=\ldots=I^5=0$ and a single nondegeneracy condition $\widetilde{X}_1\wedge \widetilde{X}_2\wedge \widetilde{X}_3\neq 0$. Solutions of $\mathcal{S}$ are then characterized by thirteen explicit algebraic equations, with eight independent ones selected dynamically along each solution.
As noted in Remark~\ref{Rem_allsolareinvar}, even solutions on the common level set that fail the nondegeneracy condition are invariant under additional symmetries.

The situation exemplified by the potential Boussinesq system, in which a finite-dimensional system $\mathcal{E}_X$ inherits a scaling symmetry that rescales sufficiently many independent first integrals, appears to be a structural feature of a class of $(1+1)$-dimensional integrable systems. In these systems, conservation laws yield hierarchies of commuting symmetries via Hamiltonian operators, and the scaling symmetry acts compatibly on the entire hierarchy. The resulting description of invariant solutions through algebraic equations determined by constants of $X$-invariant motion and the decomposition of the scaling symmetry differs qualitatively from the picture familiar for Liouville-integrable ODEs arising in Hamiltonian mechanics, where symmetries that rescale sufficiently many first integrals are not typical. It should be emphasized that the framework of Section~\ref{SectionScalsymexectsols} applies to non-integrable PDE systems as well, including overdetermined systems with more than two independent variables arising, for example, in reduction.

Several directions for further research appear natural.

The formalism of Lagrangian multiforms, mentioned in Section~\ref{Mainsection}, provides a variational perspective on integrable hierarchies in which the closure property of a Lagrangian $d$-form encodes integrability~\cite{Suris2013, SleighNijhoffCaudrelier2020}. Recent work has demonstrated that Lagrangian multiforms arise naturally as higher conservation laws of dispersionless PDEs in three dimensions and in the context of Gibbons--Tsarev equations~\cite{FerapontovVermeeren2025}, and that a Lagrangian $1$-form on coadjoint orbits captures the Poisson involutivity structure of finite-dimensional integrable systems~\cite{CaudrelierDellAttiHarland2025}. Since Theorem~\ref{Theorem1} extends directly to the spectral sequence for the Lagrangian formalism, it would be of interest to understand how the shift rule interacts with the closure property and whether the emergence of invariance at the level of internal Lagrangians (in the sense of~\cite{DRUZHKOV2023104848}) gives rise to Lagrangian multiform structures on the reduced systems.

The presymplectic BV-AKSZ approach to gauge theories~\cite{GrigorievPresympGauge2023, DneprovGrigorievGritzaenko2024} provides an invariant geometric description of Lagrangian systems beyond the jet-bundle framework. Since the invariant reduction mechanism essentially relies on the intrinsic geometry of PDEs, establishing a precise connection between presymplectic gauge PDEs and the invariant reduction of variational principles, including the case of rescaling symmetries, may yield a more unified treatment of reduction for gauge-theoretic models. In a related direction, as noted in Part~III~\cite{druzhkov2025invariantreductionpartialdifferential}, describing Poisson brackets in terms of intrinsic geometry in the general case remains a challenging open problem. Its solution would clarify the geometric origin of Poisson structures and, in combination with Theorem~\ref{Theorem1}, would enable a fully intrinsic treatment of how scaling symmetries interact with their reductions.

A practically important extension concerns conservation-law-preserving numerical discretizations. Structure-preserving schemes that respect conservation laws and symmetries of the underlying PDE are known to exhibit superior long-time accuracy and stability~\cite{CheviakovDorodnitsynKaptsov2020}. Conservation laws depending on arbitrary functions, such as the family employed in Section~\ref{LRTsubsec}, yield infinitely many constraints that a discretization can potentially inherit or approximate. An approximate conservation law framework for perturbed PDE systems~\cite{CheviakovYang2025} offers a further avenue: understanding how the shift rule of Theorem~\ref{Theorem1} behaves under perturbation could inform the design of numerical methods that preserve the emergence-of-invariance mechanism at the discrete level.

Finally, further development of the multi-reduction setting remains of interest. The approach of Anderson and Fels~\cite{AndersonFels} to the reduction of variational bicomplexes, as well as the symmetry multi-reduction method of Anco and Gandarias~\cite{AncoGandarias2020}, allow one to reduce under suitable multi-dimensional Lie groups of point symmetries. As observed in Part~III~\cite{druzhkov2025invariantreductionpartialdifferential}, a slight extension of the Anderson--Fels approach to the invariant part of the zero page of Vinogradov's $\mathcal{C}$-spectral sequence for a given equation may enable Poisson bracket reductions under non-solvable algebras of point or contact symmetries. Combining this with the shift rule of Theorem~\ref{Theorem1} may provide a direction towards a systematic theory of multi-reduction for rescaled geometric structures.

\section*{Acknowledgments }
K.D. thanks the Pacific Institute for the Mathematical Sciences for support through a PIMS
Postdoctoral Fellowship. A.C. is grateful to NSERC of Canada for support through a Discovery grant RGPIN-2024-04308.

{\footnotesize
\bibliography{references24}
\bibliographystyle{ieeetr}}

\appendix

\section{Proof of Lemma~\ref{Lemma1}}\label{App:A}

\noindent
\textbf{Proof.} At any point of $\mathcal{S}$, the vector fields $Y_i$ span the corresponding tangent plane. Hence $Y_s = f^i Y_i$ for some functions $f^i\in C^{\infty}(\mathcal{S})$.
Any two points $s_0$ and $s_1$ of a solution of $\mathcal{S}$ can be connected by a finite sequence of flows of the lifts of some vector fields from $M$ to $\mathcal{S}$. More precisely, we first lift these vector fields to $\mathcal{E}_X$ and then restrict them to $\mathcal{S}$. This sequence of flows moves a neighborhood of $s_0$ in $\mathcal{S}$ to a neighborhood of $s_1$ in $\mathcal{S}$. Any linear combination of $Y_s$, $Y_1$, $\ldots$, $Y_k$ is preserved by such flows. Then the respective coefficients of the linear combinations coincide at points of a given solution, and hence all $f^i$ are constant on solutions of $\mathcal{S}$.
Finally, from $Y_s = f^i Y_i$, \eqref{actiononcommut}, and $X|_{\mathcal{E}_X} = 0$, it follows that
\begin{align*}
-c^i_j Y_i = [Y_j, Y_s] = Y_j(f^i)Y_i\,.
\end{align*}
Then at every point of $\mathcal{S}$, one has $Y_j\, \lrcorner\, df^i = - c^i_j$ due to the linear independence of $Y_1$, $\ldots$, $Y_k$. Since the matrix $(c^i_j)$ is invertible, $df^1$, $\ldots$, $df^k$ cannot be linearly dependent at a point of $\mathcal{S}$.

%
%
%
%
%
%
%
%
%

\end{document}